\documentstyle[10pt,epsf,epsfig,dp_delphititle,cite,mcite]{dp_delphi}

\makeindex
\def\DpPaperGroup{PH--EP}
\def\DpPaperRef{2009-003}
\def\DpDate{4 February 2009}
\def\DpAuthors{DELPHI Collaboration}
\def\DpSubmit{(Accepted by Euro. Phys. J. C)}
\def\DpTitle{{Correlations between Polarisation States of $W$ Particles \\
in the
Reaction \\ $e^{-}e^{+}{\rightarrow}W^{-}W^{+}$ \\
at LEP2 Energies 189--209 GeV  }}
\def\DpComment{}
\def\DpEMail{}

\begin{document}
\makeatletter
\newcount\@tempcntc
\def\@citex[#1]#2{\if@filesw\immediate\write\@auxout{\string\citation{#2}}\fi
  \@tempcnta\z@\@tempcntb\m@ne\def\@citea{}\@cite{\@for\@citeb:=#2\do
    {\@ifundefined
       {b@\@citeb}{\@citeo\@tempcntb\m@ne\@citea\def\@citea{,}{\bf ?}\@warning
       {Citation `\@citeb' on page \thepage \space undefined}}%
    {\setbox\z@\hbox{\global\@tempcntc0\csname b@\@citeb\endcsname\relax}%
     \ifnum\@tempcntc=\z@ \@citeo\@tempcntb\m@ne
       \@citea\def\@citea{,}\hbox{\csname b@\@citeb\endcsname}%
     \else
      \advance\@tempcntb\@ne
      \ifnum\@tempcntb=\@tempcntc
      \else\advance\@tempcntb\m@ne\@citeo
      \@tempcnta\@tempcntc\@tempcntb\@tempcntc\fi\fi}}\@citeo}{#1}}
\def\@citeo{\ifnum\@tempcnta>\@tempcntb\else\@citea\def\@citea{,}%
  \ifnum\@tempcnta=\@tempcntb\the\@tempcnta\else
   {\advance\@tempcnta\@ne\ifnum\@tempcnta=\@tempcntb \else \def\@citea{--}\fi
    \advance\@tempcnta\m@ne\the\@tempcnta\@citea\the\@tempcntb}\fi\fi}
 
\makeatother
\begin{titlepage}
\pagenumbering{roman}

\CERNpreprint{\DpPaperGroup}{\DpPaperRef}   
\date{{\small\DpDate}}              
\title{\DpTitle}                
\address{\DpAuthors}                

\begin{shortabs}                
\noindent
\noindent
In a study of the reaction $e^{-} e^{+}{\rightarrow}W^{-} W^{+}$ with the DELPHI
detector, the probabilities of the two W particles occurring in the joint 
polarisation 
states transverse--transverse ($TT$), longitudinal--transverse plus transverse--longitudinal 
($LT$) 
and longitudinal--longitudinal ($LL$) have been determined using the final states 
$WW{\rightarrow}{\it l\nu  q \bar{q}}$ ~(${\it l = e,\mu}$). The two--particle 
joint polarisation probabilities, i.e. the spin density matrix elements $\rho_{TT}$,
$\rho_{LT}$, $\rho_{LL}$, 
are measured as functions of the $W^{-}$ production angle, $\theta_{W^{-}}$, at an 
average reaction energy of 198.2 GeV.  Averaged over all $\cos\theta_{W^{-}}$, 
the following joint probabilities are obtained: 

\begin{eqnarray}
\bar{\rho}_{TT} &=& (67\pm8)\% ,\nonumber\\
\bar{\rho}_{LT} &=& (30\pm8)\% ,\nonumber\\
\bar{\rho}_{LL} &=& (3\pm7)\% .\nonumber
\end{eqnarray}
\par\noindent
These results are in agreement with the Standard Model predictions of 
63.0\%, 28.9\% and 8.1\%, respectively. The related polarisation cross-sections  $\sigma_{TT}$, 
$\sigma_{LT}$ and $\sigma_{LL}$ are also presented.

\end{shortabs}

\vfill

\begin{center}
\DpSubmit \ \\      
\DpComment \ \\
\DpEMail \ \\
\end{center}

\vfill
\clearpage

\headsep 10.0pt

\addtolength{\textheight}{10mm}
\addtolength{\footskip}{-5mm}
\begingroup
%
\newcommand{\DpName}[2]{\hbox{#1$^{\ref{#2}}$},\hfill}
\newcommand{\DpNameTwo}[3]{\hbox{#1$^{\ref{#2},\ref{#3}}$},\hfill}
\newcommand{\DpNameThree}[4]{\hbox{#1$^{\ref{#2},\ref{#3},\ref{#4}}$},\hfill}
\newskip\Bigfill \Bigfill = 0pt plus 1000fill
\newcommand{\DpNameLast}[2]{\hbox{#1$^{\ref{#2}}$}\hspace{\Bigfill}}

%
\footnotesize
\noindent
\DpName{J.Abdallah}{LPNHE}
\DpName{P.Abreu}{LIP}
\DpName{W.Adam}{VIENNA}
\DpName{P.Adzic}{DEMOKRITOS}
\DpName{T.Albrecht}{KARLSRUHE}
\DpName{R.Alemany-Fernandez}{CERN}
\DpName{T.Allmendinger}{KARLSRUHE}
\DpName{P.P.Allport}{LIVERPOOL}
\DpName{U.Amaldi}{MILANO2}
\DpName{N.Amapane}{TORINO}
% ...
\normalsize
\endgroup

\newpage
\titlefoot{Department of Physics and Astronomy, Iowa State
     University, Ames IA 50011-3160, USA
    \label{AMES}}
\titlefoot{Physics Department, Universiteit Antwerpen,
     Universiteitsplein 1, B-2610 Antwerpen, Belgium
    \label{ANTWERP}}
\titlefoot{IIHE, ULB-VUB,
     Pleinlaan 2, B-1050 Brussels, Belgium
    \label{BRUSSELS}}
\titlefoot{Physics Laboratory, University of Athens, Solonos Str.
     104, GR-10680 Athens, Greece
    \label{ATHENS}}
\titlefoot{Department of Physics, University of Bergen,
     All\'egaten 55, NO-5007 Bergen, Norway
    \label{BERGEN}}
\titlefoot{Dipartimento di Fisica, Universit\`a di Bologna and INFN,
     Viale C. Berti Pichat 6/2, IT-40127 Bologna, Italy
    \label{BOLOGNA}}
\titlefoot{Centro Brasileiro de Pesquisas F\'{\i}sicas, rua Xavier Sigaud 150,
     BR-22290 Rio de Janeiro, Brazil
    \label{BRASIL-CBPF}}
\titlefoot{Inst. de F\'{\i}sica, Univ. Estadual do Rio de Janeiro,
     rua S\~{a}o Francisco Xavier 524, Rio de Janeiro, Brazil
    \label{BRASIL-IFUERJ}}
\titlefoot{Coll\`ege de France, Lab. de Physique Corpusculaire, IN2P3-CNRS,
     FR-75231 Paris Cedex 05, France
    \label{CDF}}
\titlefoot{CERN, CH-1211 Geneva 23, Switzerland
    \label{CERN}}
\titlefoot{Institut de Recherches Subatomiques, IN2P3 - CNRS/ULP - BP20,
     FR-67037 Strasbourg Cedex, France
    \label{CRN}}
\titlefoot{Now at DESY-Zeuthen, Platanenallee 6, D-15735 Zeuthen, Germany
    \label{DESY}}
\titlefoot{Institute of Nuclear Physics, N.C.S.R. Demokritos,
     P.O. Box 60228, GR-15310 Athens, Greece
    \label{DEMOKRITOS}}
\titlefoot{FZU, Inst. of Phys. of the C.A.S. High Energy Physics Division,
     Na Slovance 2, CZ-182 21, Praha 8, Czech Republic
    \label{FZU}}
\titlefoot{Dipartimento di Fisica, Universit\`a di Genova and INFN,
     Via Dodecaneso 33, IT-16146 Genova, Italy
    \label{GENOVA}}
\titlefoot{Institut des Sciences Nucl\'eaires, IN2P3-CNRS, Universit\'e
     de Grenoble 1, FR-38026 Grenoble Cedex, France
    \label{GRENOBLE}}
\titlefoot{Helsinki Institute of Physics and Department of Physical Sciences,
     P.O. Box 64, FIN-00014 University of Helsinki, 
     \indent~~Finland
    \label{HELSINKI}}
\titlefoot{Joint Institute for Nuclear Research, Dubna, Head Post
     Office, P.O. Box 79, RU-101 000 Moscow, Russian Federation
    \label{JINR}}
\titlefoot{Institut f\"ur Experimentelle Kernphysik,
     Universit\"at Karlsruhe, Postfach 6980, DE-76128 Karlsruhe,
     Germany
    \label{KARLSRUHE}}
\titlefoot{Institute of Nuclear Physics PAN,Ul. Radzikowskiego 152,
     PL-31142 Krakow, Poland
    \label{KRAKOW1}}
\titlefoot{Faculty of Physics and Nuclear Techniques, University of Mining
     and Metallurgy, PL-30055 Krakow, Poland
    \label{KRAKOW2}}
\titlefoot{LAL, Univ Paris-Sud, CNRS/IN2P3, Orsay, France
    \label{LAL}}
\titlefoot{School of Physics and Chemistry, University of Lancaster,
     Lancaster LA1 4YB, UK
    \label{LANCASTER}}
\titlefoot{LIP, IST, FCUL - Av. Elias Garcia, 14-$1^{o}$,
     PT-1000 Lisboa Codex, Portugal
    \label{LIP}}
\titlefoot{Department of Physics, University of Liverpool, P.O.
     Box 147, Liverpool L69 3BX, UK
    \label{LIVERPOOL}}
\titlefoot{Dept. of Physics and Astronomy, Kelvin Building,
     University of Glasgow, Glasgow G12 8QQ, UK
    \label{GLASGOW}}
\titlefoot{LPNHE, IN2P3-CNRS, Univ.~Paris VI et VII, Tour 33 (RdC),
     4 place Jussieu, FR-75252 Paris Cedex 05, France
    \label{LPNHE}}
\titlefoot{Department of Physics, University of Lund,
     S\"olvegatan 14, SE-223 63 Lund, Sweden
    \label{LUND}}
\titlefoot{Universit\'e Claude Bernard de Lyon, IPNL, IN2P3-CNRS,
     FR-69622 Villeurbanne Cedex, France
    \label{LYON}}
\titlefoot{Dipartimento di Fisica, Universit\`a di Milano and INFN-MILANO,
     Via Celoria 16, IT-20133 Milan, Italy
    \label{MILANO}}
\titlefoot{Dipartimento di Fisica, Univ. di Milano-Bicocca and
     INFN-MILANO, Piazza della Scienza 3, IT-20126 Milan, Italy
    \label{MILANO2}}
\titlefoot{IPNP of MFF, Charles Univ., Areal MFF,
     V Holesovickach 2, CZ-180 00, Praha 8, Czech Republic
    \label{NC}}
\titlefoot{NIKHEF, Postbus 41882, NL-1009 DB
     Amsterdam, The Netherlands
    \label{NIKHEF}}
\titlefoot{National Technical University, Physics Department,
     Zografou Campus, GR-15773 Athens, Greece
    \label{NTU-ATHENS}}
\titlefoot{Physics Department, University of Oslo, Blindern,
     NO-0316 Oslo, Norway
    \label{OSLO}}
\titlefoot{Dpto. Fisica, Univ. Oviedo, Avda. Calvo Sotelo
     s/n, ES-33007 Oviedo, Spain
    \label{OVIEDO}}
\titlefoot{Department of Physics, University of Oxford,
     Keble Road, Oxford OX1 3RH, UK
    \label{OXFORD}}
\titlefoot{Dipartimento di Fisica, Universit\`a di Padova and
     INFN, Via Marzolo 8, IT-35131 Padua, Italy
    \label{PADOVA}}
\titlefoot{Rutherford Appleton Laboratory, Chilton, Didcot
     OX11 OQX, UK
    \label{RAL}}
\titlefoot{Dipartimento di Fisica, Universit\`a di Roma II and
     INFN, Tor Vergata, IT-00173 Rome, Italy
    \label{ROMA2}}
\titlefoot{Dipartimento di Fisica, Universit\`a di Roma III and
     INFN, Via della Vasca Navale 84, IT-00146 Rome, Italy
    \label{ROMA3}}
\titlefoot{DAPNIA/Service de Physique des Particules,
     CEA-Saclay, FR-91191 Gif-sur-Yvette Cedex, France
    \label{SACLAY}}
\titlefoot{Instituto de Fisica de Cantabria (CSIC-UC), Avda.
     los Castros s/n, ES-39006 Santander, Spain
    \label{SANTANDER}}
\titlefoot{Inst. for High Energy Physics, Serpukov
     P.O. Box 35, Protvino, (Moscow Region), Russian Federation
    \label{SERPUKHOV}}
\titlefoot{J. Stefan Institute, Jamova 39, SI-1000 Ljubljana, Slovenia
    \label{SLOVENIJA1}}
\titlefoot{Laboratory for Astroparticle Physics,
     University of Nova Gorica, Kostanjeviska 16a, SI-5000 Nova Gorica, Slovenia
    \label{SLOVENIJA2}}
\titlefoot{Department of Physics, University of Ljubljana,
     SI-1000 Ljubljana, Slovenia
    \label{SLOVENIJA3}}
\titlefoot{Fysikum, Stockholm University,
     Box 6730, SE-113 85 Stockholm, Sweden
    \label{STOCKHOLM}}
\titlefoot{Dipartimento di Fisica Sperimentale, Universit\`a di
     Torino and INFN, Via P. Giuria 1, IT-10125 Turin, Italy
    \label{TORINO}}
\titlefoot{INFN,Sezione di Torino and Dipartimento di Fisica Teorica,
     Universit\`a di Torino, Via Giuria 1,
     IT-10125 Turin, Italy
    \label{TORINOTH}}
\titlefoot{Dipartimento di Fisica, Universit\`a di Trieste and
     INFN, Via A. Valerio 2, IT-34127 Trieste, Italy
    \label{TRIESTE}}
\titlefoot{Istituto di Fisica, Universit\`a di Udine and INFN,
     IT-33100 Udine, Italy
    \label{UDINE}}
\titlefoot{Univ. Federal do Rio de Janeiro, C.P. 68528
     Cidade Univ., Ilha do Fund\~ao
     BR-21945-970 Rio de Janeiro, Brazil
    \label{UFRJ}}
\titlefoot{Department of Radiation Sciences, University of
     Uppsala, P.O. Box 535, SE-751 21 Uppsala, Sweden
    \label{UPPSALA}}
\titlefoot{IFIC, Valencia-CSIC, and D.F.A.M.N., U. de Valencia,
     Avda. Dr. Moliner 50, ES-46100 Burjassot (Valencia), Spain
    \label{VALENCIA}}
\titlefoot{Institut f\"ur Hochenergiephysik, \"Osterr. Akad.
     d. Wissensch., Nikolsdorfergasse 18, AT-1050 Vienna, Austria
    \label{VIENNA}}
\titlefoot{Inst. Nuclear Studies and University of Warsaw, Ul.
     Hoza 69, PL-00681 Warsaw, Poland
    \label{WARSZAWA}}
\titlefoot{Now at University of Warwick, Coventry CV4 7AL, UK
    \label{WARWICK}}
\titlefoot{Fachbereich Physik, University of Wuppertal, Postfach
     100 127, DE-42097 Wuppertal, Germany \\
\noindent
{$^\dagger$~deceased}
    \label{WUPPERTAL}}
\nopagebreak
\clearpage

\headsep 30.0pt
\end{titlepage}

\pagenumbering{arabic}                  
\setcounter{footnote}{0}                %
\large




\section{Introduction}

\noindent
In the study of the reaction $e^{-}e^{+}{\rightarrow }W^{-}W^{+}$ at LEP2, the DELPHI 
Collaboration~\cite{DELPHIsdm} and other LEP experiments~\cite{LEPsdm1,LEPsdm2} 
have established that on 
average ${\sim}$ 25\% of $W$ particles are longitudinally polarised, as predicted by the
Standard Model. The present study  measures the joint two-particle spin density matrix elements 
which give 
the probabilities that both $W$s are transversely 
polarised ($W_{T}W_{T}$), both are 
longitudinally polarised ($W_{L}W_{L}$) or that one $W$ is transversely polarised while the 
associated $W$ is longitudinal ($W_{T}W_{L}$+$W_{L}W_{T}$). In what follows, these correlations 
will be referred to as $TT$, $LL$ and $LT$,  respectively.
This is a more detailed test of the Standard Model 
prediction for the $W$ polarisation than those previously published.  Production of  longitudinal 
$W$s is of particular interest because they are associated with the breaking of the electroweak 
symmetry. This study tests the theoretical prediction of the correlations and in particular, 
that the correlation $LL$ is suppressed relative to $LT$.

\vspace{3mm}
\noindent
The previously published measurements of the spin-dependent correlations between the $W$ 
particles in the reaction  $e^{-}e^{+}{\rightarrow}W^{-}W^{+}$ are by the OPAL collaboration \cite{Opal}
and by the L3 collaboration~\cite{L3}. OPAL evaluated the cross-sections $\sigma_{TT}$, $\sigma_{LT}$ 
and $\sigma_{LL}$ from their data at 189 GeV with a comparatively low statistics. Their results are in 
poor agreement with the Standard Model.
L3 used  the $WW{\rightarrow} l\nu q \bar{q}$ and 
$WW {\rightarrow} q \bar{q}~ q \bar{q}$ events from their full LEP2 data set to establish the 
correlation between the decay planes of the two $W$s. The correlation was found to be consistent 
with the Standard Model prediction.

\vspace{3mm}
\noindent
The analysis presented in this  paper uses only the events in which one $W$ decays into an 
electron plus a neutrino or a muon plus neutrino, while the other $W$ decays into two hadron jets.
These ``semi--leptonic" events are kinematically well constrained and they offer the best 
available data for any detailed analysis of this reaction. The $\tau$ semi--leptonic events are 
excluded because the uncertainties in their identification cause problems in $WW$ correlation 
measurements. The fully hadronic final state $WW{\rightarrow}q\bar{q}q\bar{q}$ 
is also excluded because of the uncertainties in  jet reconstruction: the charges of the hadron 
jets cannot be well measured and the particles from the four jets tend to overlap in the space of the
detector,  resulting in uncertainties in associations between the $W$s and the measured jets.

\vspace{3mm}
\noindent
A complete description of the polarisation states of the produced $W$ particles is given in terms 
of the two-particle joint spin density matrix 
$\rho_{\lambda_{1}{\lambda_{1}}^{\prime}\lambda_{2}{\lambda_{2}}^{\prime}}$, where $\lambda_{1}$ 
and $\lambda_{2}$ are the helicities of the $W^{-}$ and $ W^{+}$ respectively. In terms of the $W$ production 
helicity amplitudes,  $F^{(\mu)}_{\lambda_{1}\lambda_{2}}$,  the spin density matrix elements are defined by
$$\rho_{\lambda_{1}{\lambda_{1}}^{\prime}\lambda_{2}{\lambda_{2}}^{\prime}} \equiv 
\frac{\sum_{\mu}{F^{(\mu)}_{\lambda_{1}\lambda_{2}}}{F^{(\mu)\ast}_{{\lambda_{1}}^{\prime}{\lambda_{2}}^{\prime}}}}
{\sum_{\mu\lambda_{1}\lambda_{2}}|{F^{(\mu)}_{\lambda_{1}\lambda_{2}}}|^{2}}.$$
\noindent
The normalisation  is such that the trace of the matrix is unity. The initial state helicity sum runs over
$\mu = \pm1/2$ and the $W$ particle helicities run over $\lambda_{i},{\lambda_{i}}^{'} = \pm1, 0$.

\vspace{3mm}
\noindent
The helicities of $W$ particles can be determined from their centre-of-mass decay distribution asymmetries.
The above definition of the $\rho_{\lambda_{1}{\lambda_{1}}^{\prime}\lambda_{2}{\lambda_{2}}^{\prime}}$
elements can be put \cite{Bilenky} into the following form which is model independent and is directly 
applicable to experimental data corrected for backgrounds and detection efficiences: 
 
\newpage 
\begin{eqnarray}
\frac{d\sigma}{d(\cos\theta_{W^{-}})d(\cos\theta^{*}_{1})d\phi^{*}_{1}d(\cos\theta^{*}_{2})d
\phi^{*}_{2}}= \hspace{3.0in}\\
\frac{d\sigma}{d(\cos\theta_{W^{-}})}\left(\frac{3}{8\pi}\right)^{2}
\sum_{\lambda_{1}{\lambda_{1}}^{\prime}\lambda_{2}{\lambda_{2}}^{\prime}}
\rho_{\lambda_{1}{\lambda_{1}}^{\prime}\lambda_{2}{\lambda_{2}}^{\prime}}
(\cos\theta_{W^{-}}) D_{\lambda_{1}{\lambda_{1}}^{\prime}}(\theta^{*}_{1},\phi^{*}_{1}) 
D_{\lambda_2 {\lambda_{2}}^{\prime}}(\theta^{*}_{2},\phi^{*}_{2}) \,. \nonumber
\end{eqnarray}

\vspace{3mm}
\noindent
Here, $\theta_{W^{-}}$ is the angle of the $W^{-}$ production with respect to the $e^{-}$ beam, 
$\theta^{*}_{1}$ and $\theta^{*}_{2}$ are the polar decay angles of the  $W^{-}$ and $W^{+}$  in 
their rest frames and $\phi^{*}_{1}$, $\phi^{*}_{2}$ are the corresponding azimuthal decay angles, as shown in 
figure 5.
The functions $D_{\lambda\lambda^{\prime}}$ are the theoretical decay distributions of the $W$ 
particles in the helicity states specified by the  $\lambda$ indices. Precise definitions of the 
angles $\theta^{*}_{1,2}$ and $\phi^{*}_{1,2}$ and of the functions $D_{\lambda\lambda^{\prime}}$ 
relevant to the present analysis are  given in section~3. It should be noted that the cross-section 
formula (1) is model-independent regarding the $WW$ production process.

\vspace{3mm}
\noindent
The single-$W$ particle spin density matrix $\rho_{\lambda\lambda^{\prime}}$, derived from the 
$WW$ spin density matrix 
$\rho_{\lambda_{1}{\lambda_{1}}^{\prime}\lambda_{2}{\lambda_{2}}^{\prime}}$ by summation over one 
of the indices ($1,2$), gives 
information about the polarisaton of one $W$ regardless of the state of the other.  All nine 
elements of the single-$W$ particle spin density matrix can be determined using the data from 
semi-leptonic events. This was done in~\cite{DELPHIsdm} and~\cite{LEPsdm1,LEPsdm2}, where only 
the electron 
and muon decays of one $W$ were used as the analyser of the $W$ polarisation. The hadronic decays 
of $W$ particles were not used  because of the reasons outlined above and because the analysing 
power of polarisation in such decays is greatly reduced as the result of the severe practical 
difficulty to distinguish  quark jets from  anti-quark jets. 

\vspace{3mm}
\noindent
In the present study, it would ideally be desirable to  measure the complete $9\times9$ 
matrix $\rho_{\lambda_{1}{\lambda_{1}}^{\prime}\lambda_{2}{\lambda_{2}}^{\prime}}$. As the result 
of the limitations in the polarisation information from the hadronic $W$ decays, only a small part 
of the joint $WW$ spin density matrix can be measured. It is possible to  measure 5 (out of 9) 
diagonal elements ($\rho_{\lambda_{1}\lambda_{1}\lambda_{2}\lambda_{2}}$) plus 9 complex 
off--diagonal elements.  Instead of this incomplete set of individual matrix elements, the 
following three linear combinations of matrix elements are considered in this paper:
\begin{eqnarray}
\rho_{LL} &=& \rho_{0000} \, ,\nonumber\\
\rho_{TT}&=&\rho_{++--}+\rho_{--++}+\rho_{----}+\rho_{++++}\  ,\\ 
\rho_{LT} &=& \rho_{++00}+\rho_{00++}+\rho_{--00}+\rho_{00--}\  .\nonumber
\end{eqnarray}

\noindent
The quantities $\rho_{LL}$, $\rho_{TT}$ and $\rho_{LT}$ are composed of the diagonal elements of 
the full matrix and they can be interpreted as probabilities of the joint, i.e. correlated, 
polarisation states of the two $W$s.  The elements $\rho_{TT}$ and $\rho_{LT}$ do not distinguish 
between the + and -- transverse helicities, and also the polarisations of the states $W^{-}_{L} 
W^{+}_{T}$  and  $W^{-}_{T} W^{+}_{L}$ are combined. This is a reduced set of information about 
the $WW$ polarisations but it is nevertheless very useful. The elements $\rho_{TT}$, $\rho_{LT}$ 
and $\rho_{LL}$ can  be measured in  semi-leptonic $WW$ events because cancellations  in the 
sums~(2) imply that 
the incompleteness of the  polarisation information in the hadronic $W$ decays does not 
matter~\cite{Bilenky}. In section 3 it will be shown how $\rho_{TT}$, $\rho_{LT}$ and $\rho_{LL}$ 
can be measured directly from the data without recourse to the individual spin density matrix 
elements in~(2).
 
\section{The Experiment, Treatment of Data and Simulation}
\subsection{The Experiment}
\noindent
The DELPHI detector is described in detail in~\cite{DELSIM1,DELSIM2}  and its configuration during the LEP2  
runs is given in~\cite{DELPHIww}. The reference \cite{DELPHIww}  gives a complete description 
of the selection of $WW$ events in DELPHI. The present analysis uses the data taken at  
centre-of-mass energies between 189 and 209~GeV. The data are grouped into three sets at average 
energies of 189~GeV, 200~GeV and 206~GeV. The total integrated luminosity is 520~$\mathrm{pb^{-1}}$, and 
the luminosity--weighted average energy of all data is~198.2 GeV.
Jet reconstruction algorithms as well as electron and muon identification are also described  in 
~\cite{DELPHIww}.

\subsection{Selection of Data and Monte Carlo Simulation}
\label{sec:selection}
\subsubsection{ Data selection}
\noindent
The initial selection procedure for the channels  $\mu$$\nu${\it q}$\bar{q}$  and $e\nu${\it 
q}$\bar{q}$ is based on the typical topology of those events. As already mentioned, events from 
the $\tau$$\nu${\it q}$\bar{q}$ channels are not included in this analysis and thus they are a 
part of the background. The starting values of the data cuts are those listed in~\cite{DELPHIww}:

\vspace{3mm}
(i) Visible event energy $\geq 40\%$ of the nominal centre-of-mass energy;
\par
(ii) The event transverse energy $\geq 45$ GeV;
\par
(iii) The event must have at least one muon or one electron identified;
\par
(iv) Electron or muon candidate's momentum $\geq 20$~GeV/$c$;
\par
(v) Charged lepton track angle with respect to the beam direction $\geq 20^{\circ}$;
\par
(vi) The total track multiplicity in each hadron jet $\geq 3$;
\par
(vii) Reconstructed $ W^{-}$ and $ W^{+}$ masses  $\geq 50$~GeV/$c^2$.\footnote{The 
masses are calculated from the results of a preliminary 1-constraint  kinematic 
fit to the reaction $e^{-}e^{+}{\rightarrow}W^{-}W^{+}.$}
\par

\vspace{3mm}
\noindent
The precise values of these cuts, in particular those on the event transverse energy and the minimum
particle multiplicity in jets, were varied slightly for data taken at the three different average 
$e^{-}e^{+}$ 
energies.  Three-constraint kinematic fits were then performed to the reaction 
$e^{-}e^{+}{\rightarrow} W^{-}W^{+}$ on the selected data samples, requiring both $W$s to have
the same mass consistent with  80.4~GeV/$c^2$. Cuts on the $\chi^{2}$ probability distribution were 
then applied, with the 
value of the cut (in the region 0.5\% --1\%) determined from the $\chi^{2}$ distribution in each 
of the three data sets. The final sample, taken at all beam energies, consists of 800 electron plus 
880 muon events. This sample is somewhat smaller than that reported in  \cite{DELPHIww}  because we 
require  full functionality of all parts of DELPHI.
\par
\vspace{3mm}
\noindent
Particle momenta and angles obtained from kinematic fitting have been used in the analysis of this 
experiment.

\subsubsection{Event Simulation}

\begin{figure}[h]
\begin{center}
\epsfig{file=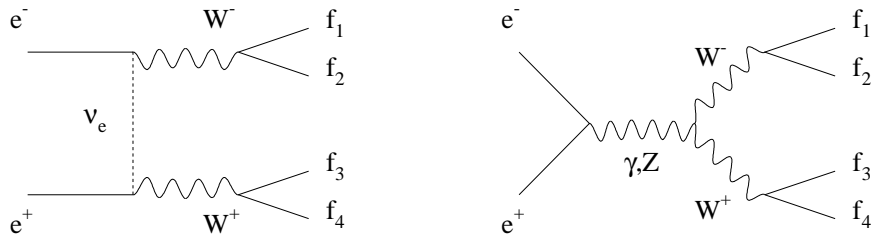}
\caption{CC03 diagrams, i.e. the lowest order contributions to the amplitude for $W^{-}W^{+}$ 
production. ($\rm f_{1,2,3,4}$ stand for the appropriate fermions.)}
\label{fig:cc03}  
\end{center}
\end{figure}

\noindent
Simulation of events plays a crucial role in the experimental procedure to separate  events 
corresponding to the reaction $e^{-}e^{+}{\rightarrow}W^{-}W^{+}$, the ``signal",  from 
backgrounds. The signal is defined by the three CC03 Feynman diagrams shown in 
figure~\ref{fig:cc03}. These account only for a part of the four-fermion processes contributing to 
the data. In this experiment  a customised version~\cite{WPHACT1}  of the WPHACT~\cite{WPHACT2,WPHACT3} 
generator program was used to simulate all the four-fermion processes. The DELPHI WPHACT program 
includes reweighting for the Double Pole Approximation (DPA) radiative corrections and the 
possibility to
compute the matrix elements of different subsets of Feynman diagrams. The weights are the ratios 
of the squared matrix element for $WW$ production only via the CC03 diagrams to that for 
production via the full set of four-fermion processes.  It is thus possible to simulate CC03 
events corresponding to production via the CC03 diagrams with or without inclusion of other 
four-fermion processes.

\vspace{3mm}
\noindent
In addition to the four-fermion background, there is a significant two-fermion background, mostly 
from $\bar{q}q\gamma$ final states. This background has been simulated using the KK2F 
generator~\cite{KK}.

\vspace{3mm}
\noindent
The generators were interfaced to the PYTHIA~\cite{PYTHIA1,PYTHIA2,PYTHIA3}  hadronisation 
program. Large simulated 
samples, of the order of $10^{6}$ events,  were produced by the programs listed above, interfaced 
to the DELPHI detector simulation program DELSIM~\cite{DELSIM1,DELSIM2}. 

\subsection{Reconstruction of events}
\label{sec:reconstruction}
\noindent
Event reconstruction efficiencies were determined using events simulated with the WPHACT Monte 
Carlo program adapted for 
DELPHI~\cite{WPHACT1}. The efficiency is defined as the ratio of the number of reconstructed CC03 
WPHACT Monte Carlo events to the number of generated events, including all four-fermion channels.   

\vspace{3mm}
\noindent
Efficiencies are determined as functions of $\cos\theta_{W^{-}}$, $\cos\theta^{*}_{1}$ and 
$\cos\theta^{*}_{2}$, defined in section 1. The resulting reconstruction efficiency table is a  
$5\times5\times7$ matrix corresponding to 7 bins of $\cos\theta_{W^{-}}$, as defined in section 4.2, 
and 5 equal size bins for each 
of $\cos\theta^{*}_{1}$ and $\cos\theta^{*}_{2}$. The reconstruction efficiency matrices were 
determined separately for the electron 
and muon event samples at the three centre-of-mass energies. Typical values of the 
reconstruction efficiency, $\epsilon$, as a function of $\cos\theta_{W^{-}}$  are 0.65 to 0.74 
when integrated over all $\cos\theta^{*}_{1}$ and $\cos\theta^{*}_{2}$. The values of $\epsilon$ 
as a function of $\cos\theta^{*}_{1}$ (or of $\cos\theta^{*}_{2}$), averaged over 
$\cos\theta_{W^{-}}$, vary over the range 0.6 to 0.8.

\subsection{Treatment of Backgrounds}
\label{sec:backgrounds}
\noindent
The dominant backgrounds to the selection of $e \nu jj$ and $\mu \nu jj$ events (where $j$ implies 
a hadronic jet) can be divided into two groups:

\vspace{3mm}
(a)  {\it Events which result from problems in reconstruction or selection procedures.}  The 
dominant contribution to this class of events comes from two-fermion final states, in particular 
from $\bar{q}q\gamma$. Other contributions come from neutral current four-fermion final states 
which might be misidentified as $e \nu jj$ or, more rarely, as $\mu \nu jj$. Misidentified charged 
current events  $\tau \nu jj$  are also potentially a background in this experiment. 

\vspace{3mm}
(b)  {\it The non-CC03 charged current four-fermion contributions to the global $\mu \nu jj$ or $e 
\nu jj$  final states.}  In contrast to the background (a), this background would exist even in a 
perfect detector and an ideal experimental event selection procedure. A number of the charged 
current four-fermion diagrams in this category can interfere with the CC03 amplitude\footnote{As a 
consequence of the finite $W$ width, any Monte Carlo generator of $e^{-}e^{+}{\rightarrow} 
W^{-}W^{+}$ events at low energy must be a four-fermion generator in order to satisfy gauge 
invariance.}.

\vspace{3mm}
\noindent
Following the treatment in~\cite{DELPHIww}, the backgrounds requiring special attention are:

\vspace{3mm}
{\bf(i)} the $\tau$ events, $WW{\rightarrow}\tau \nu jj$ (belonging to group (a));
\par
{\bf (ii)} various four-fermion processes  which for experimental or other reasons lead to the 
same final state as the reaction $e^{-}e^{+}{\rightarrow} W^{-}W^{+}$  (such backgrounds may arise 
from groups (a) and (b));
\par
{\bf(iii)} two--fermion events, mostly $\bar{q}q\gamma$ interactions (group (a)). 

\noindent
Discussions of each of these background sources follows in turn.

\vspace{3mm}
\noindent
$\bullet$ The background from source (i) has been investigated by passing the simulated $\tau \nu 
jj$ events through the normal analysis chain and requiring $e \nu q \bar{q}$ or $\mu\nu q \bar{q}$ 
fits. The background from this source turns out to be negligible as the result of kinematic cuts 
and fitting.
\par

\vspace{3mm}
\noindent
$\bullet$ The background (ii) is due both to charged current and neutral current events.

\vspace{3mm}
\noindent
The charged current processes consist of the three CC03 diagrams plus seven diagrams with 
$s$-channel exchange of  $Z^{0}/\gamma$, leading to the production of only one $W$. There are also 
ten charged current diagrams corresponding to $t$-channel processes which 
give rise to one $W$. The latter diagrams can give rise to backgrounds only in the electron final 
state $e\nu jj$.

\vspace{3mm}
\noindent 
The neutral current four-fermion states have two quarks and two leptons of the same flavour. If one
of the electrons is not identified in the detector, the event may be classified as belonging to 
the channel $WW \rightarrow e \nu jj$ and may satisfy the criteria for an acceptable 
kinematic fit. Altogether, the four-fermion backgrounds affect the electron channel, 
$e \nu q \bar{q}$, more than the muon channel $\mu\nu q \bar{q}$.

\begin{figure}[h!]
\begin{center}
\epsfig{file=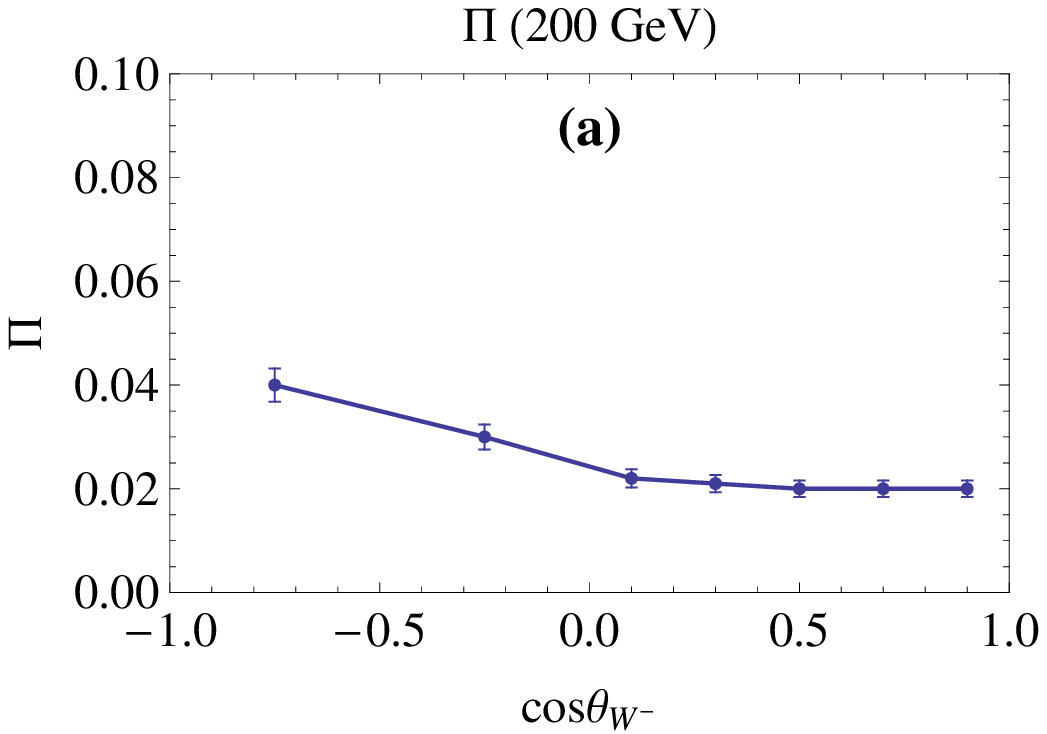}  
\epsfig{file=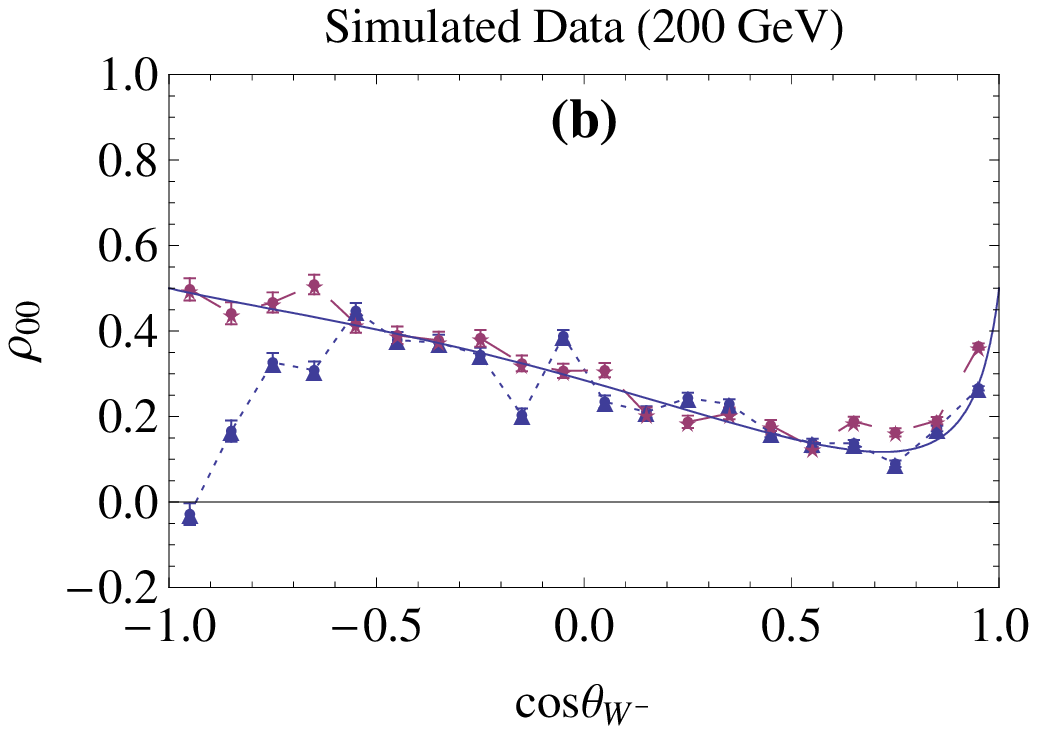}\\
\caption{ (a) Plot of the quantity $\Pi$, defined in the text, as a function of $\cos\theta_{W^-}$ 
for data simulated at 200~GeV. 
\newline
\hspace*{1.70cm}
(b) Plots 
of the density matrix element $\rho_{00}$ evaluated from four-fermion events generated at 200~GeV 
with WPHACT before (triangle symbols) and after (star symbols) the event reconstruction procedure, 
including application of data cuts and kinematic fitting. The smooth curve is from an analytic calculation 
using CC03 diagrams.}
\label{fig:4f1}
\end{center}
\end{figure}

\vspace{3mm}
\noindent
It was found that the non-CC03  four-fermion background in the real events could be efficiently 
reduced to a level less than 4\% by tuning the 
kinematic cuts and the $\chi^{2}$ cuts, as described in section~\ref{sec:selection}. The 
effectiveness of the removal of this class of background events can be demonstrated in the 
following way: Starting from a large sample of generated events, two data samples were 
produced making use of the event reweighting facilities in WPHACT. Sample A contained predominantly 
the CC03 events and sample B the non-CC03 four-fermion background. Each sample was processed through
the  experimental procedure described in section 2.2.   
The event ratio $\Pi={\rm B/A}$ represents the proportion of the four-fermion background in the $WW$ signal 
to be expected in the final sample of the real data. This quantity is of the order of 3\%  and is 
weakly dependent on $\cos\theta_{W^{-}}$. A plot of $\Pi$ for WPHACT data at 200~GeV is shown in 
figure~\ref{fig:4f1}(a). The results for other run energies are similar.  
   
\vspace{3mm} 
\noindent
A further test of the effectiveness of the method for dealing with the non-CC03 four-fermion 
background is to apply the analysis to a simulated data set where the expected result is known.
For this purpose,  the element $\rho_{00}$ of the single-$W$ spin density matrix 
$\rho_{\lambda \lambda^{\prime}}$ is evaluated. This test was carried out for data generated at 
all three run energies, but here only the results from WPHACT at 200~GeV  are shown. In 
figure~\ref{fig:4f1}(b), triangle symbols are used to plot the value of $\rho_{00}$ obtained from 
the 
WPHACT generated events using all the four-fermion diagrams,  with no cuts except for the final 
kinematic $\chi^{2}$ selection. Star symbols are used to plot the value of $\rho_{00}$ after the 
same generated events have been passed through the complete selection procedure described in 
section~\ref{sec:selection}.
The smooth solid curve is from an analytic calculation using only the CC03 diagrams. The 
conclusion drawn from this is that the procedure adopted for analysing the data removes 
essentially all the four-fermion background, leaving events which are attributable to the CC03 
signal.\footnote{The plot made with triangle symbols in Fig.\ref{fig:4f1}(b) shows a strong dip 
near $\cos\theta_{W^-}= -1$ and some fluctuations up to about $\cos\theta_{W^-}= 0$ . These effects  
are caused by some of the  four-fermion backgrounds in the simulated raw data. In particular, 
the background events whose W particles 
are not genuine spin 1 states tend to interfere destructively with the true Ws in  the 
$\rho_{00}$ evaluation.}

\vspace{3mm}
\noindent
The residue of the non-$WW$ four-fermion background in the real data at all run energies is 
estimated to be at the level of $3\pm 2\%$, the uncertainty is due to combining  results from 
different energies.

\begin{figure}[h!]
\begin{center}
\epsfig{file=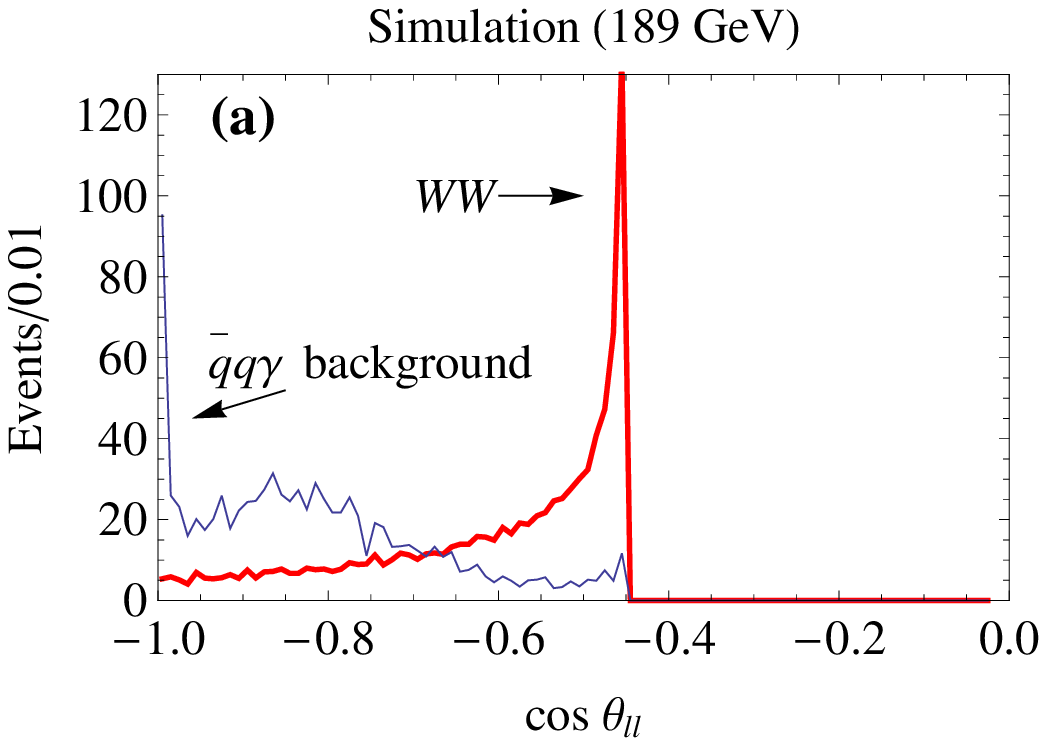}
\epsfig{file=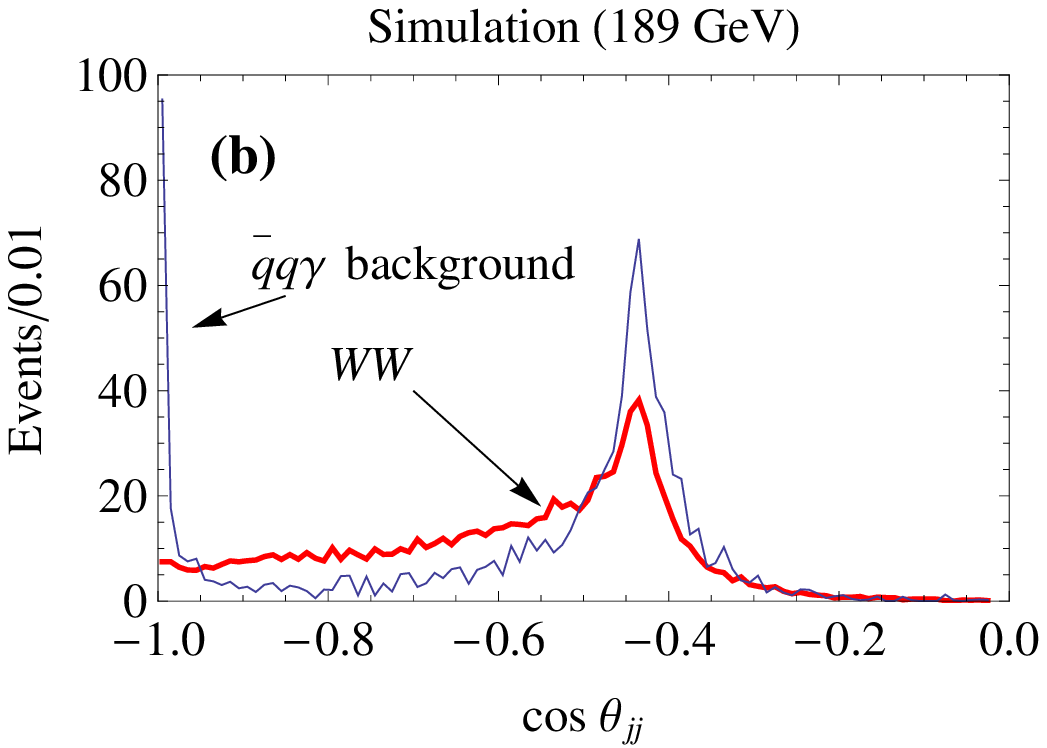}
\caption{Distributions (a) of $\cos\Theta_{ll}$ and (b) of $\cos\Theta_{jj}$ in the $\bar{q}q 
\gamma$ background and in the reaction $e^{-}e^{+}{\rightarrow}W^{-}W^{+}$ from data simulated at 
189~GeV. (The angles $\Theta_{ll}$  and $\Theta_{jj}$ are defined in the text.) The
curves labelled $\bar{q}q \gamma$  and $WW$ are normalised to the same number of events.}
\label{fig:qqg1}
\end{center}
\end{figure}

\begin{figure}[h!]
\begin{center}
\epsfig{file=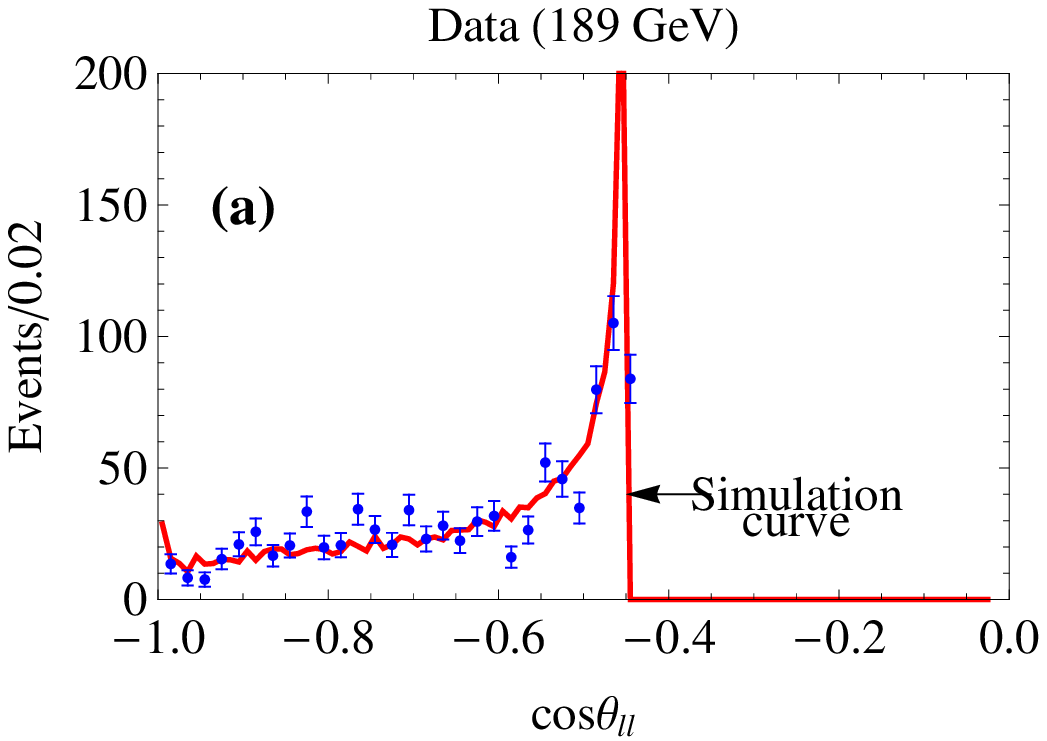}
\epsfig{file=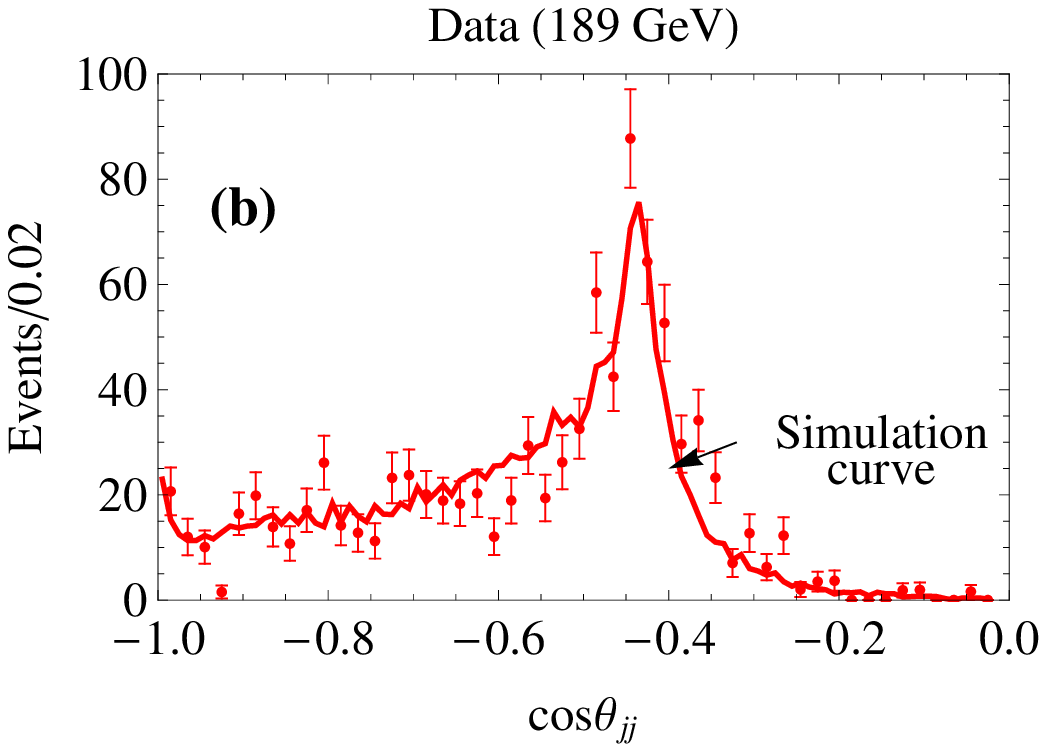}
\caption{ Distributions (a) of $\cos\Theta_{ll}$ and (b) of $\cos\Theta_{jj}$ for the real data at 
189~GeV. The superimposed curves are the result
of a fit using simulated samples of  $\bar{q}q \gamma$ and $WW$ events.}
\label{fig:qqg2}
\end{center}
\end{figure}

\vspace{3mm}
\noindent
$\bullet$ The background process (iii) is potentially very serious because the cross-section  for
$\bar{q}q \gamma$ production is about an order of magnitude larger than that for the $WW$ signal.  
Although the  topology of the $ \bar{q}q \gamma$ events is quite different from that of the $WW$ 
events, the reconstructed events of the background can resemble and fit the $WW$ 
reaction. The problem of how to suppress this background has been investigated  using the KK2F 
Monte Carlo generator~\cite{KK}. It is found that this background shows a very characteristic 
kinematic signature in the distribution of the quantities $\cos\Theta_{ll}$ and $\cos\Theta_{jj}$, 
where  $\Theta_{ll}$ is the angle between the momentum vectors of the two leptons (one charged and 
a  neutrino) coming from one 
$W$ and  $\Theta_{jj}$ is the angle between the jets from the accompanying $W$, with quantities 
defined in the laboratory frame (i.e. the reaction centre-of-mass).\footnote{The momentum vectors 
are 
taken from the constrained kinematic fit.} The presence of the $\bar{q}q \gamma$ background shows 
up as accumulations of events at
$\cos\Theta_{ll} \sim -1$ and  $\cos\Theta_{jj} \sim -1$. 

\vspace{3mm}
\noindent
After the usual cuts and kinematic fitting, the real data samples show peaks in the 
$\cos\Theta_{ll}$ and $\cos\Theta_{jj}$ distributions indicating small but non-negligible 
contamination from the $\bar{q}q \gamma$ background. Below it will be shown that the 
contamination is of the order of 10\% at 189~GeV. This contamination has to be evaluated 
accurately because it affects the angular distribution of $W$ decays. The kinematics of the $W$ decay 
in the laboratory frame of the reaction is such that the distributions of the quantities 
$\cos\Theta_{ll}$ and $\cos\Theta_{jj}$ have large discriminating power against the $\bar{q}q\gamma$ 
background. This enables the magnitude of the background to be determined 
and also provides a means of reducing the background by applying cuts on these 
distributions or by applying suitable event weights. Only the method of weights has 
been used in dealing with this background.

\vspace{3mm}
\noindent
Figures~\ref{fig:qqg1}(a) and~\ref{fig:qqg1}(b) show, respectively, simulations of the 
$\cos\Theta_{ll}$ and $\cos\Theta_{jj}$ distributions at 189~GeV. The curves labelled $WW$ 
were obtained from the WPHACT simulation of CC03, while the 
curves labelled $\bar{q}q \gamma$ are from the KK2F simulation of this background. At this stage 
of the analysis, the relative magnitudes of the $WW$ and $ \bar{q}q \gamma$ components are still 
unknown; therefore, the two curves are normalised to the same number of events. 
Figures~\ref{fig:qqg2}(a) and~\ref{fig:qqg2}(b) show the results of a least squares fit to the 
relative contributions from $WW$ and $ \bar{q}q \gamma$, using the distributions from the 
simulated events, to the real data distributions. The fit to the 189~GeV data requires 
contributions of  90\% from $WW$ and $10\pm 2\%$ from $\bar{q} q\gamma$, while the same analysis applied 
to the data at 200~GeV and 206~GeV require a $5\pm 2\%$ $\bar{q} q \gamma$ background contribution at 
both energies..

\vspace{3mm}
\noindent
This determination of the relative magnitude of the $\bar{q}q \gamma$ background with respect to 
the $WW$  signal enables a combined WPHACT  plus KK2F simulation of the real data. 
The combined simulation was subsequently used  to derive weighting factors, $p$, defining the 
purity of the $WW$ signal in each bin of $\cos\theta_{W^{-}}$, $\cos\theta^{*}_{1}$ and 
$\cos\theta^{*}_{2}$, using the same binning as that used for the reconstruction efficiencies, 
$\epsilon$. The application of the weight factors $p$ is described in the following section.


\par
\section{Analysis}
\label{sec:analysis}

\begin{figure}[h!]
\begin{center}
\epsfig{file=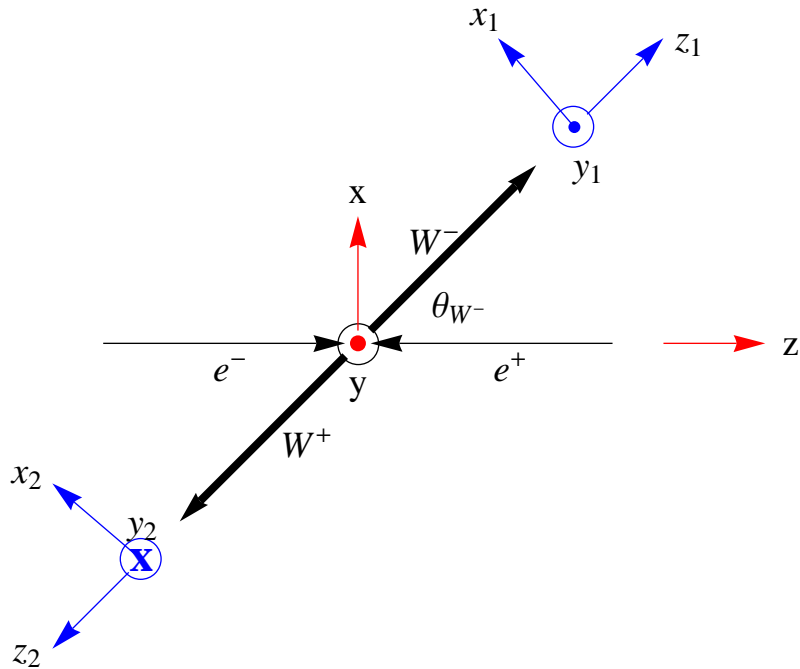}
\caption{Diagram of the momentum vectors in the reaction plane of the process 
$e^{-}e^{+}{\rightarrow} W^{-}W^{+}$.  The two right-handed sets of orthogonal axes 
($x_{1}$,$y_{1}$,$z_{1}$)  and ($x_{2}$,$y_{2}$,$z_{2}$) in the respective rest frames of the 
$W^{-}$ and $W^{+}$ are also shown. The polar angles $\theta^{*}_{1}$ and $\theta^{*}_{2}$ of the 
vector momenta of the two-body decays of the $W^{-}$ and $W^{+}$ are measured with respect to the 
axes $z_{1}$ and $z_{2}$.} 
\label{fig:mypict1}
\end{center}
\end{figure}

\noindent
$W$ decays are well described by the V--A theory of the charged current weak interactions. The 
theory gives the functions $D_{\lambda \lambda^{\prime}}(\theta^{*},\phi^{*})$ required for 
evaluating  the spin density matrix elements  
$\rho_{\lambda_{1}{\lambda_{1}}^{\prime}\lambda_{2}{\lambda_{2}}^{\prime}}$ by applying equation 
(1) to the data. The $W^-$ production angle, $\theta_{W^-}$, and the $W^{-,+}$ decay angles 
$\theta_{1,2}^{*}$, $\phi_{1,2}^{*}$, which specify the direction of the final state fermion in 
the rest frame of the $W^{-}$ and of the final state anti--fermion in the rest frame of the 
$W^{+}$, are defined in figure~\ref{fig:mypict1}. 

\noindent
As already pointed out, fermions can be distinguished from anti-fermions in the $W$ leptonic 
decays  but not so in the hadronic decays. However, some of the decay functions $D_{\lambda 
\lambda^{\prime}}(\theta^{*},\phi^{*})$ are invariant  under the transformation which rotates the 
vector momentum of a fermion in the $W$ rest frame into the direction of its opposite anti-fermion 
vector. These functions will be called symmetric and designated by $D^{(s)}$. Consequently, the 
symmetric $D$ functions and the symmetric parts of the non-symmetric $D$ 
functions can be used to analyse the polarisation of the $W$s decaying into the purely hadronic 
final states. The polarisation information obtained is thus  incomplete but, nevertheless, it is 
useful and, in particular,  the quantities  $\rho_{TT}$,  $\rho_{LT}$ and $\rho_{LL}$ can be 
obtained from the data using only the symmetric decay distributions in both associated $W$s.

\vspace{3mm} 
\noindent
The theoretical formalism for extracting $\rho_{TT}$, $\rho_{LT}$ and $\rho_{LL}$ from the data is 
based on equation (1) with two modifications:

\vspace{3mm}
(i)  Equation (1) is integrated over the full range of $\phi^{*}_{1}$ and  $\phi^{*}_{2}$. This 
removes the functions $D_{\lambda\lambda^{\prime}}$ having $\lambda^{\prime}\neq\lambda$ and 
eliminates all non-diagonal elements of the matrix  
$\rho_{\lambda_{1}\lambda_{1}^{\prime}\lambda_{2}\lambda_{2}^{\prime}}$. The following three decay 
functions  remain: 

\begin{eqnarray}
D_{++}(\theta^{*}) & = & \frac{1}{2}(1-\cos\theta^{*})^{2},\nonumber\\
D_{00}(\theta^{*}) & = &  \sin^{2} \theta^{*} ,\\
D_{--}(\theta^{*}) & = & \frac{1}{2}(1+\cos\theta^{*})^{2}.\nonumber
\end{eqnarray}

(ii)  Furthermore, only the symmetric parts, $D^{(s)}_{\lambda\lambda}$,  of the functions in (3) 
are to be used:

\begin{eqnarray}
D^{(s)}_{++}(\theta^*) = D^{(s)}_{--}(\theta^*) = 
\frac{1}{2}(1 + \cos^{2}\theta^{*}),\nonumber\\
D^{(s)}_{00}(\theta^*) = \sin^{2}\theta^{*}.\hspace{1.5cm}\nonumber\\
\nonumber
\end{eqnarray}

\noindent
Pursuing the above steps, the following modified  form~\cite{Bilenky} of equation (1) is 
obtained:

\begin{eqnarray}
\frac{d\sigma^{(s,s)}}{d(\cos\theta_{W^{-}})d(\cos\theta^{*}_{1})d(\cos\theta^{*}_{2})}= 
\left( \frac{3}{4} \right)^{2}\left(\frac{d\sigma}{d(\cos\theta_{W^{-}})}\right)
\{\rho_{TT}(\cos\theta_{W^{-}})D^{(s)}_{++}(\theta^{*}_{1})D^{(s)}_{++}(\theta^{*}_{2})
\nonumber\\
+\rho_{LT}(\cos\theta_{W^{-}})[D^{(s)}_{00}(\theta^{*}_{1})D^{(s)}_{++}(\theta^{*}_{2})
\nonumber
+ D^{(s)}_{++}(\theta^{*}_{1})D^{(s)}_{00}(\theta^{*}_{2})]\nonumber
+\rho_{LL}(\cos\theta_{W^{-}})D^{(s)}_{00}(\theta^{*}_{1})D^{(s)}_{00}(\theta^{*}_{2})\} \, ,\\
\end{eqnarray}

\noindent
where $\rho_{TT}$, $\rho_{LL}$ and $\rho_{LT}$ have been defined in (2). The above equation 
assumes $CP$ invariance, i.e. that  the 
joint polarisations  $ W^{-}_{L} W^{+}_{T}$  and  $W^{-}_{T} W^{+}_{L}$ are equal. The superscript 
$(s,s)$ indicates that only the symmetric decays of both $W$s are considered.

\begin{figure}[h]
\begin{center}
\epsfig{file=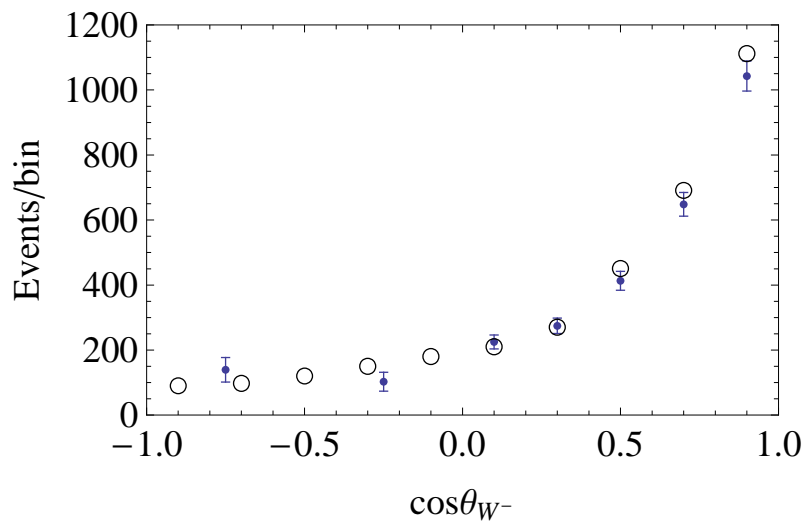}
\caption{ Angular distribution of $W^{-}$ production. Points with error bars are the experimental 
data; the errors shown are statistical only. The data binning is explained in section 4.2. The small 
circles show the predictions of the WPHACT simulation.}
\label{fig:angdist}  
\end{center}
\end{figure}

\vspace{3mm}
\noindent
The angular distribution of the $W^{-}$ production is not used explicitly in the data analysis 
for measuring the correlations $\rho_{TT}$, $\rho_{LT}$ and $\rho_{LL}$.  
However, it is appropriate to examine  the shape of the distribution 
$dN/d(\cos\theta_{W^{-}})$ as a test of the quality of the data. Figure~\ref{fig:angdist} shows 
the angular distribution of the $W^{-}$ in the present data at all energies combined, compared to 
the WPHACT prediction.  The agreement with WPHACT is satisfactory. The distribution shown also 
agrees well with that in the DELPHI publication on the $WW$ production 
cross-section~\cite{DELPHIww}.

\vspace{3mm}
\noindent
Before using the functions~(3) to determine the intensities of different helicity states in the 
data, these functions must be transformed to a related set of functions with the property that 
each one is non--orthogonal to only one in the group of distribution functions (3) and  is 
orthogonal to the other two. These functions are called projectors~\cite{Gounaris} and they can 
easily be worked out from~(3). The projector functions  needed to extract $\rho_{TT}$, $\rho_{LT}$ 
and $\rho_{LL}$ are:
\begin{eqnarray}
\Lambda_{L} &=& 2 - 5\cos^{2}\theta^{*} ,\\
\Lambda_{T} &=& 5\cos^{2}\theta^{*}-1 .\nonumber
\end{eqnarray}

\par\noindent
These projectors are normalised to give the spin density matrices in the standard 
representation~\cite{Bilenky,Gounaris}. The quantities  $\rho_{LL}$, $\rho_{TT}$ and $\rho_{LT}$ 
are obtained from the data by evaluating the following sums:

\begin{eqnarray}
\rho_{LL} &=&\frac{1}{N_{w}}\sum_{i}\Lambda_{L}(\theta^{*}_{1i})w_{i}\Lambda_{L}(\theta^{*}_{2i}) 
,\nonumber\\
\rho_{TT} &=& \frac{1}{N_{w}}\sum_{i}\Lambda_{T}(\theta^{*}_{1i})w_{i}\Lambda_{T}(\theta^{*}_{2i}) 
,\\
\rho_{LT} &=& 
\frac{1}{N_{w}}\left(\sum_{i}\Lambda_{L}(\theta^{*}_{1i})w_{i}\Lambda_{T}(\theta^{*}_{2i})+\sum_{i}
\Lambda_{T}(\theta^{*}_{1i})w_{i}\Lambda_{L}
(\theta^{*}_{2i})\right) ,\nonumber
\end{eqnarray}

\par\noindent
where summations are over all events $i$. $N_{w}=\sum_{i}w_{i}$ is the sum of all event weights 
$w_{i}$, and 
the event weights $w$ are defined as $$w = \frac{p}{\epsilon},$$ where $p$ is the $WW$ purity 
factor, defined as in section~\ref{sec:backgrounds}, for each bin of $(\cos\theta_{W^-}, 
\cos\theta_{1}^{*}, \cos\theta_{2}^{*})$, and $\epsilon$ is the event 
reconstruction efficiency in that bin, defined in section~\ref{sec:reconstruction}. In these 
formulae, index 1 refers to $W^{-}$ and index 2 refers to $W^{+}$.

\vspace{3mm}
\par\noindent
Monte Carlo studies  have shown that the correlations $\rho_{LT}$, $\rho_{LL}$, $\rho_{TT}$ 
extracted from small samples of data (such as we have at each of the three energy points 
considered) are subject to large statistical fluctuations. These fluctuations are much larger than 
those encountered in the determination of the single-W spin density matrix elements. Because of 
that, all 1680 semi-leptonic electron and muon events  have been taken as one 
sample for measuring $\rho_{LT}$, $\rho_{LL}$ and $\rho_{TT}$. The sum of weights of these 
events is 2844.

\vspace{3mm}
\noindent
The fact that events which we analyse here as one sample come from a spread of centre-of-mass
energies presents no difficulty because the theoretical predictions which we want to test 
can be modified to take into account the spread. 
In particular,  the Standard Model \cite{Bilenky} predicts a negligible variation of $\rho_{TT}$, 
$\rho_{LT}$ and $\rho_{LL}$ over the energy range of this experiment.

\section{Results}
\subsection{Systematic Effects and Errors}
\par

\noindent
(a) {\it Data cuts}.
\par
\noindent
Systematic 
effects resulting from residual backgrounds in real events have been estimated by processing the data 
several times with small variations in the cuts and, separately,
with various $\chi^{2}$ cuts. Variations in the resulting values of the spin density matrix 
elements $\rho_{LT}$, $\rho_{LL}$, $\rho_{TT}$  amount to 10\% of their statistical error. This is 
interpreted as the magnitude of the systematic uncertainty and it is neglected.

\par\vspace{3mm}
\noindent
(b) {\it Hadron jet reconstruction}. 
\par
\noindent
Problems in hadron jet reconstruction \cite{DELPHIww} can  give rise to a shift in  $\cos\theta_{W^{-}}$ with the 
further consequence that the reconstruction efficiency is read from the wrong cell of the 
$\epsilon$ matrix. This migration and its effects have been examined using simulated events.
The resulting uncertainty on $\theta_{W^{-}}$ is  small: it varies from $\pm 2^{\circ}$ at 
large angles to $\pm 5^{\circ}$ at small angles, i.e. in the forward direction\footnote{This 
uncertainty includes a problem in the simulation of charged particle tracks in the forward region of 
DELPHI.  This problem and its solution are discussed in two recent DELPHI papers~\cite{forward1,forward2}.}. 
This is negligible by comparison with the sizes of the $\cos\theta_{W^{-}}$ bins. The effect of 
this migration on the spin density matrix elements has been examined by moving the
simulated events randomly by the above uncertainty in $\cos\theta_{W^{-}}$. The effect of this 
variation on the joint spin density matrix elements is  5\% -- 8\% of the statistical uncertainty 
and is therefore considered to be negligible.
\par\vspace{3mm}
\noindent
Also, jet reconstruction problems can produce wrong momentum vectors of the $W$ hadronic decay 
products. This has been investigated by processing the same events using different jet algorithms 
as described in reference \cite{DELPHIww}. No statistically significant effect was found when
comparing the spin density matrix elements distributions obtained in these tests.

\par\vspace{3mm}
\noindent 
(c) {\it Lepton charge determination}. 
\par
\noindent
Tracks at small angle with respect to the $e^{-}$ and $e^{+}$ beams are susceptible to wrong 
charge determination. This problem is 
essentially eliminated \cite{DELPHIsdm} by the $20^{\circ}$ cut  on the lepton polar angle (see 
section~\ref{sec:selection}). The effect of this cut on the spin density matrix elements has been 
examined by simulation of events and was found to be negligible.

\vspace{3mm}
\noindent
(d) {\it Radiative corrections}.
\par
\noindent
The effect of the initial state radiation is essentially removed by the appropriate energy cut. The final 
state radiative corrections are implemented through the Double Pole Approximation in the WPHACT 
reweighting. The uncertainty due to the radiative corrections on the spin density matrix elements 
is negligible. This has been established by a comparison of the spin density matrix  elements 
evaluated from events generated with WPHACT, including the corrections, with the same elements 
calculated analytically~\cite{Bilenky,Gounaris} without the radiative corrections. This can be 
understood because the spin density matrix elements are ratios of quantities which are similarly 
affected by the radiative effects.

\vspace{3mm}
\noindent
(e) {\it Use of a fixed W mass in kinematic fitting of events}.
\par
\noindent
Three-constraint kinematic fits of the reaction events are needed in order to separate the signal from
a large background. Fixing the masses of both $W$s in the reaction to the same value, as stated in section 
2.2.1,  does not cause noticeable distortions of angular distributions and other quantities needed for the 
physics analysis. This has been checked by comparing the results obtained from the three-constraint 
fits with those from the one-constraint fits of the same events. As a further test, somewhat different 
fixed values of the W mass were tried. Statistically insignificant differences in the results were found.     

\subsection{Presentation of Results}

\begin{figure}[h!]
\begin{center}
\epsfig{file=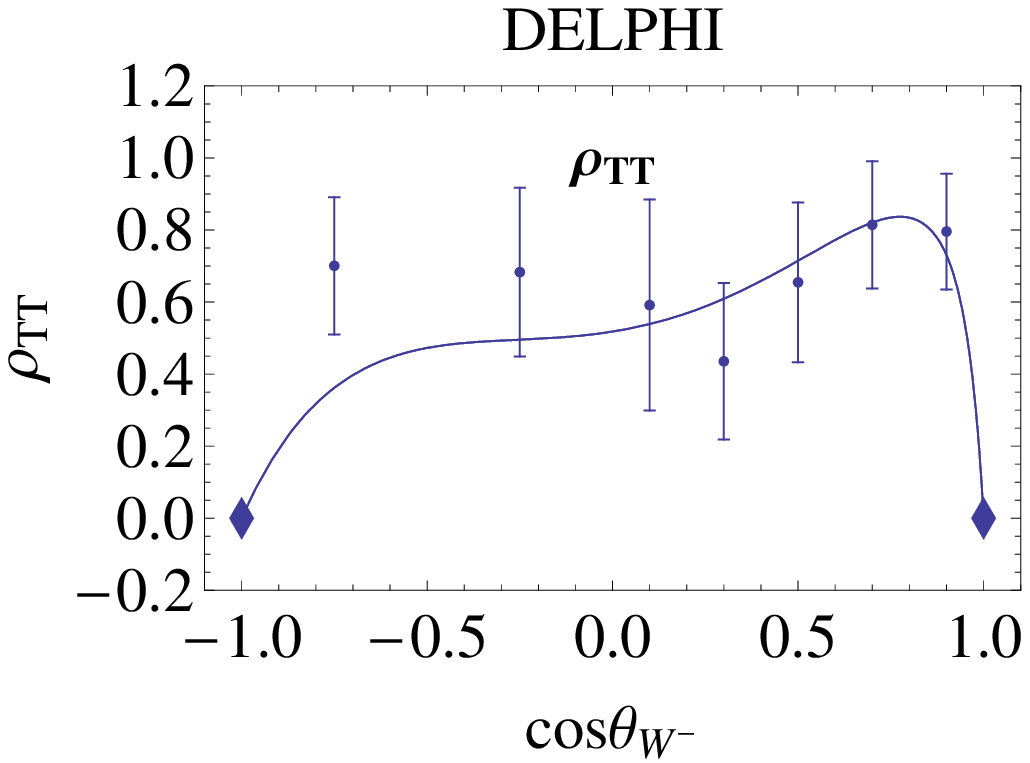,width=0.5\textwidth}\\
\vspace{-0.1cm}
\epsfig{file=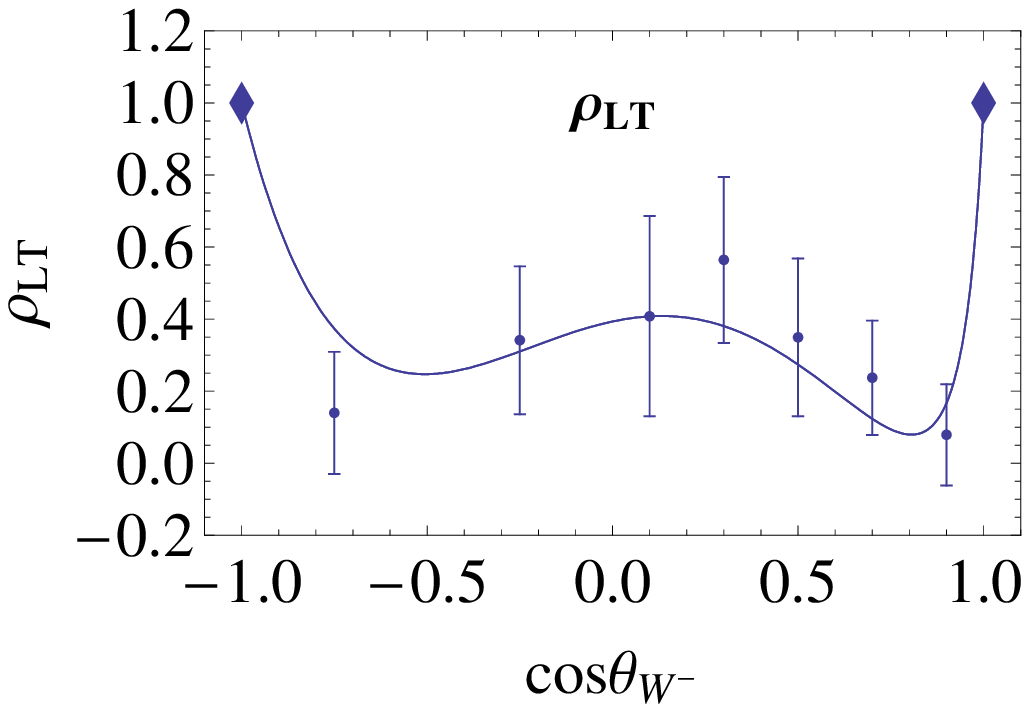,width=0.5\textwidth}\\
\vspace{-0.1cm}
\epsfig{file=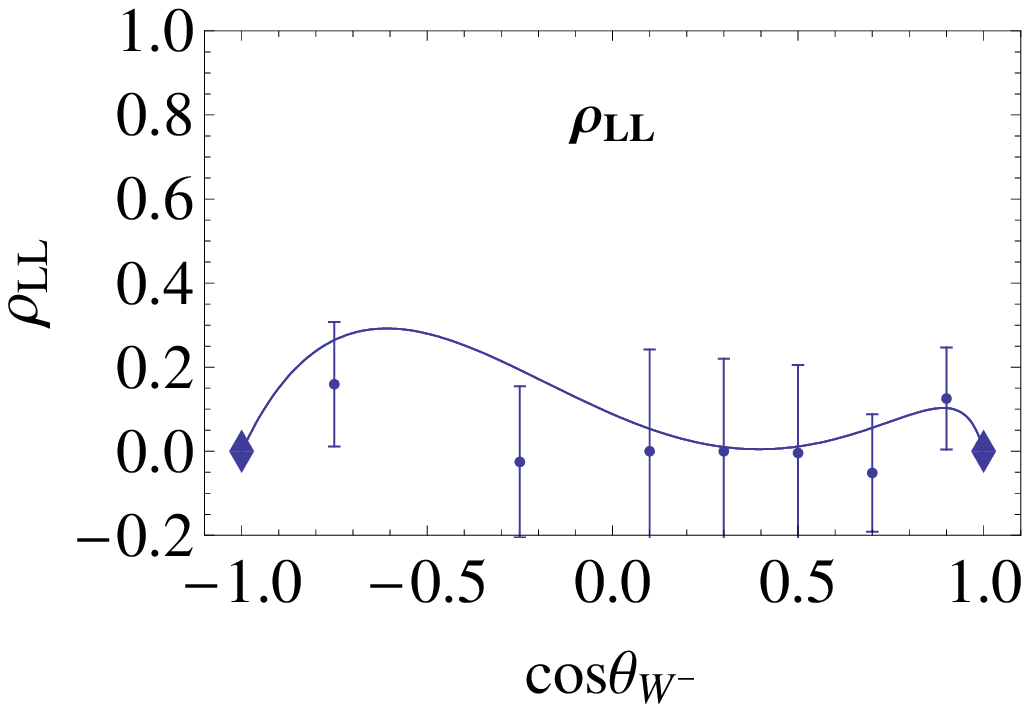,width=0.5\textwidth}\\
\caption{Variation of the spin density matrix elements $TT$, $LT$ and $LL$  with the cosine of the 
$W^{-}$ production angle, $\theta_{W^{-}}$, at the average energy of~198.2 GeV. Points with error 
bars are the measured data, and the curves are the Standard Model CC03 predictions. The diamond 
symbols indicate the points predicted by angular momentum conservation, as explained in the text 
in section~4.2.}
\label{fig:rhos}
\end{center}
\end{figure}

\noindent
The measured values of $\rho_{TT}$, $\rho_{LT}$ and $\rho_{LL}$ as functions of 
$\cos\theta_{W^{-}}$  are shown in figure~\ref{fig:rhos}. 
Because the number of events in the negative hemisphere of $ W^{-}$ production is much smaller 
than that in the positive hemisphere, 
the data have been divided into two bins in the negative hemisphere and five bins in the positive 
hemisphere. (The positive hemisphere is in the direction of the $e^{-}$ beam.) It is easy to derive
from formulas (5) and (6) that the condition 
\begin{equation}
\rho_{LL}+\rho_{TT}+\rho_{LT}=1 
\label{eqn:unitarity}
\end{equation}
\noindent
is valid on an event by event basis and is hence automatically satisfied by all data samples.

\vspace{2mm}
\noindent
The curves shown in the plots are the Standard Model calculations based on the CC03 diagrams 
evaluated at 198.2~GeV using the expressions in~\cite{Gounaris}. The error bars shown in the plots 
are statistical, the systematic errors being negligible by comparison. Errors 
on all measured quantities are evaluated from the data as standard deviations. The distributions 
of the errors are approximately Gaussian.

\vspace{3mm}
\noindent
Some events can contribute negative numbers to the sums shown in~(6) while the final result is 
positive.  However, if a particular correlation quantity $\rho$ is very small, the measurement 
errors, which have a Gaussian distribution, can lead to an overall result which is negative. This 
happens in three out of the seven measured values of $\rho_{LL}$ presented here.  They are very 
small negative quantities, consistent with zero within the measurement errors, i.e. 
$\mid\rho_{LL}\mid < \delta\rho_{LL}$, where $\delta\rho_{LL}$ is the measurement error. These 
negative values of $\rho_{LL}$ are included in figure~\ref{fig:rhos}. The 
condition~(\ref{eqn:unitarity})  still holds in these cases as a  result of the properties 
of the projector functions~(5). Of course, the physical quantities $\rho$ should satisfy the 
condition $\rho\geq0$. Several methods of adjusting the results to satisfy the positivity 
condition were tried, but not used in the end because they introduce biases that are more harmful 
than the small negative numbers among the results. 


\vspace{3mm}
\noindent
The data bins in figure~\ref{fig:rhos} are too wide to show the possibly rapid variations of the measured 
quantities very near to $\cos\theta_{W^{-}}=\pm1$. However, the exact values of these correlations 
at $\cos\theta_{W_{-}}=\pm1$ follow from the conservation 
of angular momentum. Neglecting the electron mass, the vector and axial--vector interactions 
involved at the $e^{-}$ and $e^{+}$ vertices cause the initial $e^{-}$ and $e^{+}$ to interact 
only when their helicities are opposite. Thus, the initial system has total helicity $\pm1$ and 
when the final state  is collinear with the $e^{-}e^{+}$ beams its total helicity must be the 
same. This means that at $\theta_{W^{-}} = 0$ or $\pi$, $\rho_{LT}$ must be 1 and, at the same 
time, $\rho_{LL}=\rho_{TT}=0$. These values have been indicated with diamond-shaped  
symbols in figure~\ref{fig:rhos}.

\vspace{3mm}
\noindent
Because of the low statistics in this experiment, it is useful to
examine the values of $\rho_{TT}$, $\rho_{LT}$ and $\rho_{LL}$  averaged over all bins of 
$\cos\theta_{W^{-}}$ and to compare them  with the predictions of the Standard Model. The results are 
shown in Table 1.

\begin{table}[h]
\begin{center}
\begin{tabular}{|c| |c| |c|}\hline
$\bar{\rho}$ & Measured fraction & Standard Model  \\ \hline \hline
$\bar{\rho}_{TT}$ & 67 $\pm$ 8\% & 63.0\% \\ \hline
$\bar{\rho}_{LT}$ & 30 $\pm$ 8\% & 28.9\% \\ \hline
$\bar{\rho}_{LL}$ & 3 $\pm$ 7\% & 8.1\% \\ \hline
\hline
\end{tabular}
\caption{ Measured values of $\rho_{TT}$,\ $\rho_{LT}$\  and\ $\rho_{LL}$, averaged over 
$\cos\theta_{W^{-}}$, 
compared with the predictions of the Standard Model. Errors are statistical.}
\end{center}
\end{table}

\begin{figure}[h!]
\begin{center}
\epsfig{file=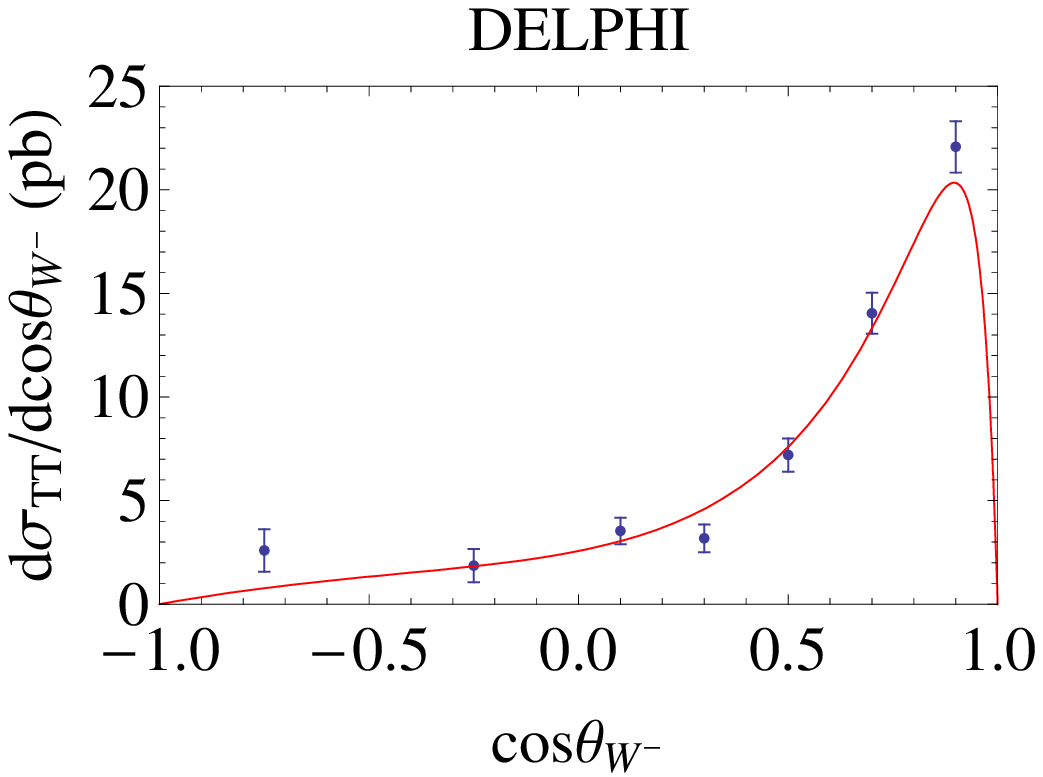,width=0.5\textwidth}\\
\vspace{-0.1cm}
\epsfig{file=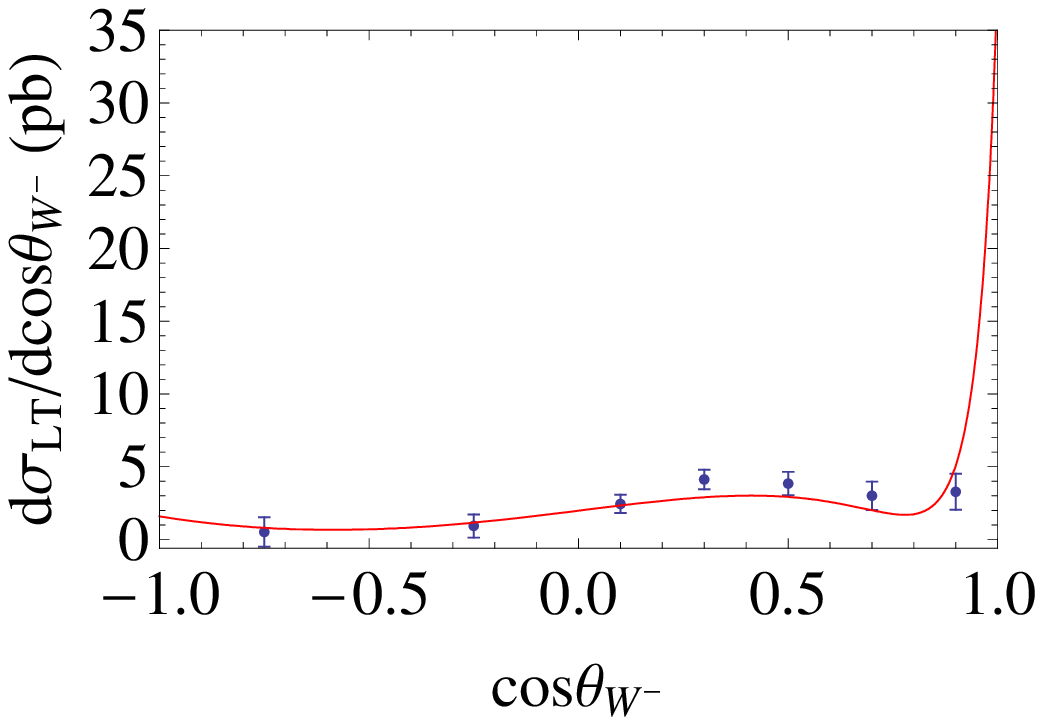,width=0.5\textwidth}\\
\vspace{-0.1cm}
\epsfig{file=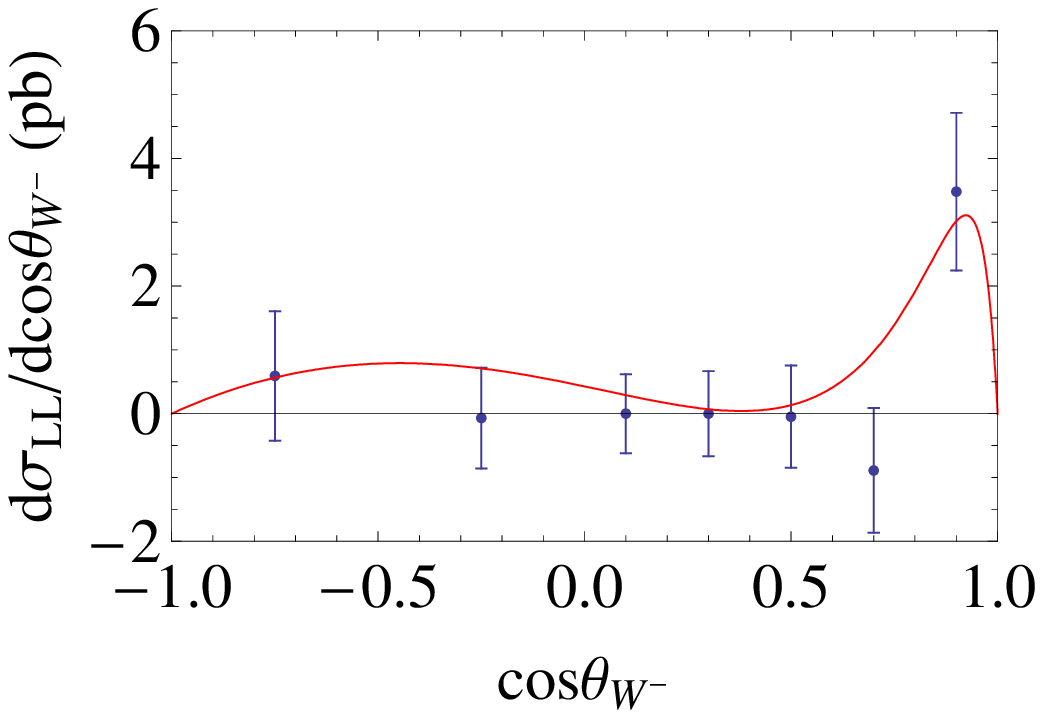,width=0.5\textwidth}\\
\caption{Variation of the differential cross-sections $TT$, $LT$ and $LL$ as functions of the 
cosine of the 
$W^{-}$ production angle, $\theta_{W^{-}}$. Points with error 
bars are the measured data;  the curves are the Standard Model CC03 predictions for the average 
energy of 198.2 GeV.} 
\label{fig:sigmas}
\end{center}
\end{figure}

\vspace{3mm}
\noindent
The differential polarisation cross-sections $d\sigma_{TT}/d(\cos\theta_{W^{-}})$,  
$d\sigma_{LT}/d(\cos\theta_{W^{-}})$ and $d\sigma_{LL}/d(\cos\theta_{W^{-}})$ are related to 
$\rho_{TT}$, $\rho_{LT}$,  $\rho_{LL}$ through the equation 

\begin {equation}
\frac{d\sigma_{TT}(\cos\theta_{W^{-}})}{d(\cos\theta_{W^{-}})}=
\frac{d\sigma(\cos\theta_{W^{-}})}{d(\cos\theta_{W^{-}})}
\rho_{TT}(\cos\theta_{W^{-}}),
\label{eqn:sigma}
\end{equation}

\noindent
plus the analogous expressions for the states LT and LL. The first term on the right-hand side of 
(\ref{eqn:sigma}) can
be replaced by the data points from the angular distribution  shown in figure 6,  normalised 
to the total cross-section for $e^{-}e^{+}{\rightarrow}W^{-}W^{+}$ at the average energy of the experiment.
The relevant measured cross-section is $\sigma = (17.07 \pm 0.57)$ pb at 198.2 GeV. It has been obtained by an
interpolation of the DELPHI measurements \cite{DELPHIww} which cover the range 161 to 209~GeV.
With this result and the
measured  $\rho_{TT}$, $\rho_{LT}$, $\rho_{LL}$ as functions of $\cos\theta_{W^{-}}$, one obtains 
the differential polarisation cross-sections shown in figure \ref{fig:sigmas}. 
Integration over  $\cos\theta_{W^{-}}$   yields the  total polarisation cross-sections  
$\sigma_{TT}$,  $\sigma_{LT}$  and  $\sigma_{LL}$ shown in Table 2.  

\vspace{3mm}
\noindent
These results are in good agreement 
with the Standard Model predictions. The polarisation fractions expressed in terms of the density
matrix elements (Table 1) and those expressed in terms of the cross-section ratios (Table 2) are 
different expressions of the two-particle polarisation correlations in the reaction.  Measurement errors 
of the polarisations are  themselves strongly correlated. The extent of all the correlations 
will be shown in figure 9(b).

\vspace{3mm}
\noindent

\begin{table}[h]
\begin{center}
\begin{tabular}{|c| |c| |c|}\hline
$\sigma$ & Measured Cross--Section & Standard Model  \\ \hline \hline
$\sigma_{TT}$ & (12 $\pm$ 1) pb & (10.57 $\pm$ 0.05) pb \\ \hline
$\sigma_{LT}$ & ( 4 $\pm$ 1) pb  & (4.95 $\pm$ 0.02) pb \\ \hline
$\sigma_{LL}$ & ( 1 $\pm$ 1) pb  & (1.40 $\pm$ 0.01) pb \\ \hline
\hline
$\sigma Ratios$ & Measured Ratio & Standard Model  \\ \hline \hline
${\sigma_{TT}}/{\sigma}$ & 0.70 $\pm$ 0.06 & 0.625 $\pm$ 0.003 \\ \hline
${\sigma_{LT}}/{\sigma}$ & 0.23 $\pm$ 0.06 & 0.292 $\pm$ 0.001 \\ \hline
${\sigma_{LL}}/{\sigma}$ & 0.06 $\pm$ 0.06 & 0.083 $\pm$ 0.001 \\ \hline
\hline
\end{tabular}
\caption{ Measured values of the total cross-sections $\sigma_{TT}$,\ $\sigma_{LT}$\ 
and $\sigma_{LL}$, 
at the average energy of 198.2~GeV, compared with the predictions of the Standard Model. Also:
Measured values of the ratios  ${\sigma_{TT}}/{\sigma}$,   ${\sigma_{LT}}/{\sigma}$
and ${\sigma_{LL}}/{\sigma}$, 
at the average energy of 198.2~GeV, compared with the predictions of the Standard Model.}
\end{center}
\end{table}


\section{Discussion and Conclusions}
\noindent
The  measurements of the total cross-section for the reaction $e^{-}e^{+}{\rightarrow} W^{-}W^{+}$ 
by ALEPH, DELPHI, L3 and OPAL~\cite{LEPewwg}, have successfully tested the predicted gauge 
coupling cancellations in this reaction. The direct measurement of the longitudinal cross-section 
for the  {\it single} (i.e. uncorrelated) $W$ production~\cite{DELPHIsdm,LEPsdm1,LEPsdm2} 
is an important test 
of the same cancellation because the cause of the potential divergence of the cross-section is in 
the longitudinal parts. The measurement of the {\it joint} $WW$ helicity states presented in this 
paper is an advance towards a more complete test of the Standard Model in the context of the $WW$ 
reaction. Our measurement of $\sigma(e^{-}e^{+}{\rightarrow}W_{L}W_{L})$  is consistent  both with 
zero and with the small value predicted  by the Standard Model.  This confirms that the 
probability of longitudinal $W$ production is predominantly in association with a transverse $W$. 

\begin{figure}[h]
\begin{center}
\begin{tabular}{l r}
\epsfig{file=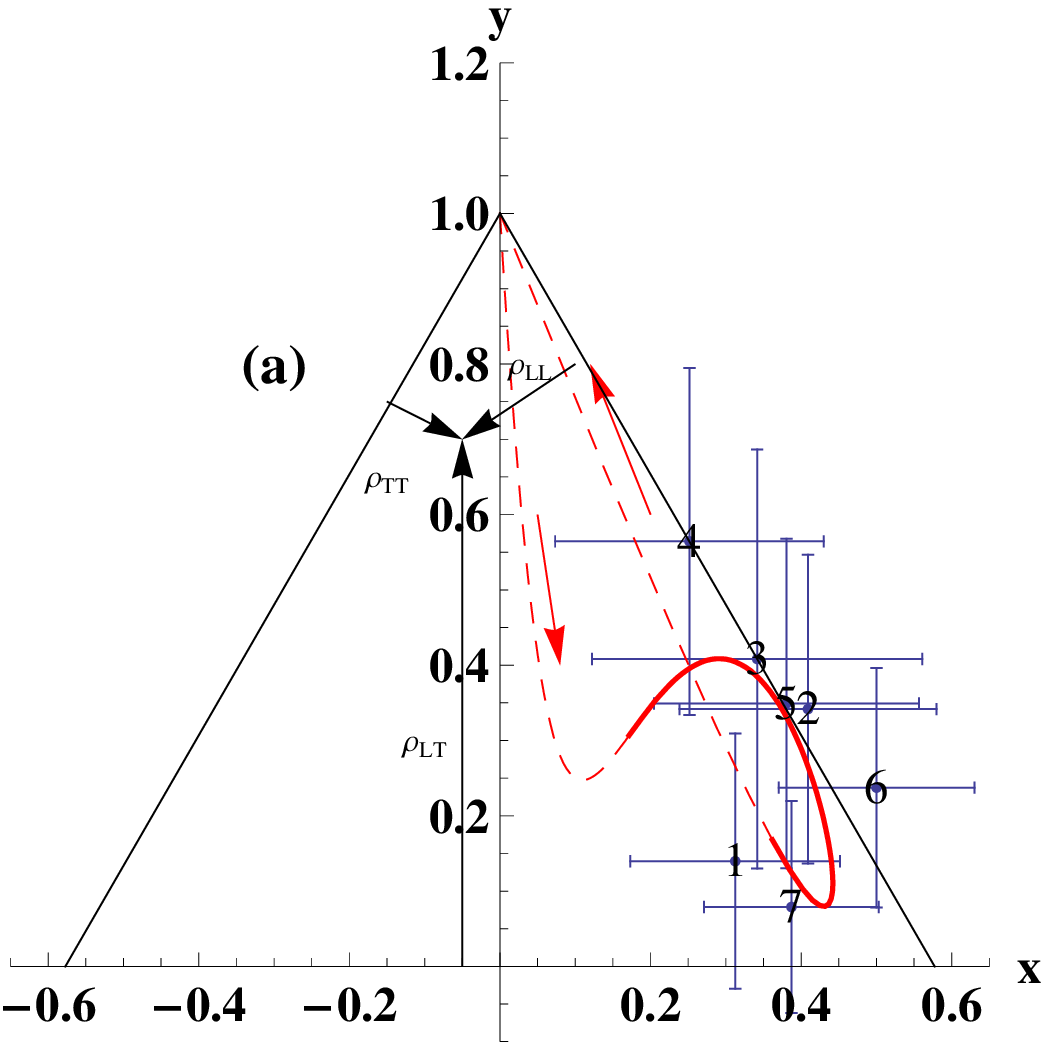,width=0.5\textwidth}&
\epsfig{file=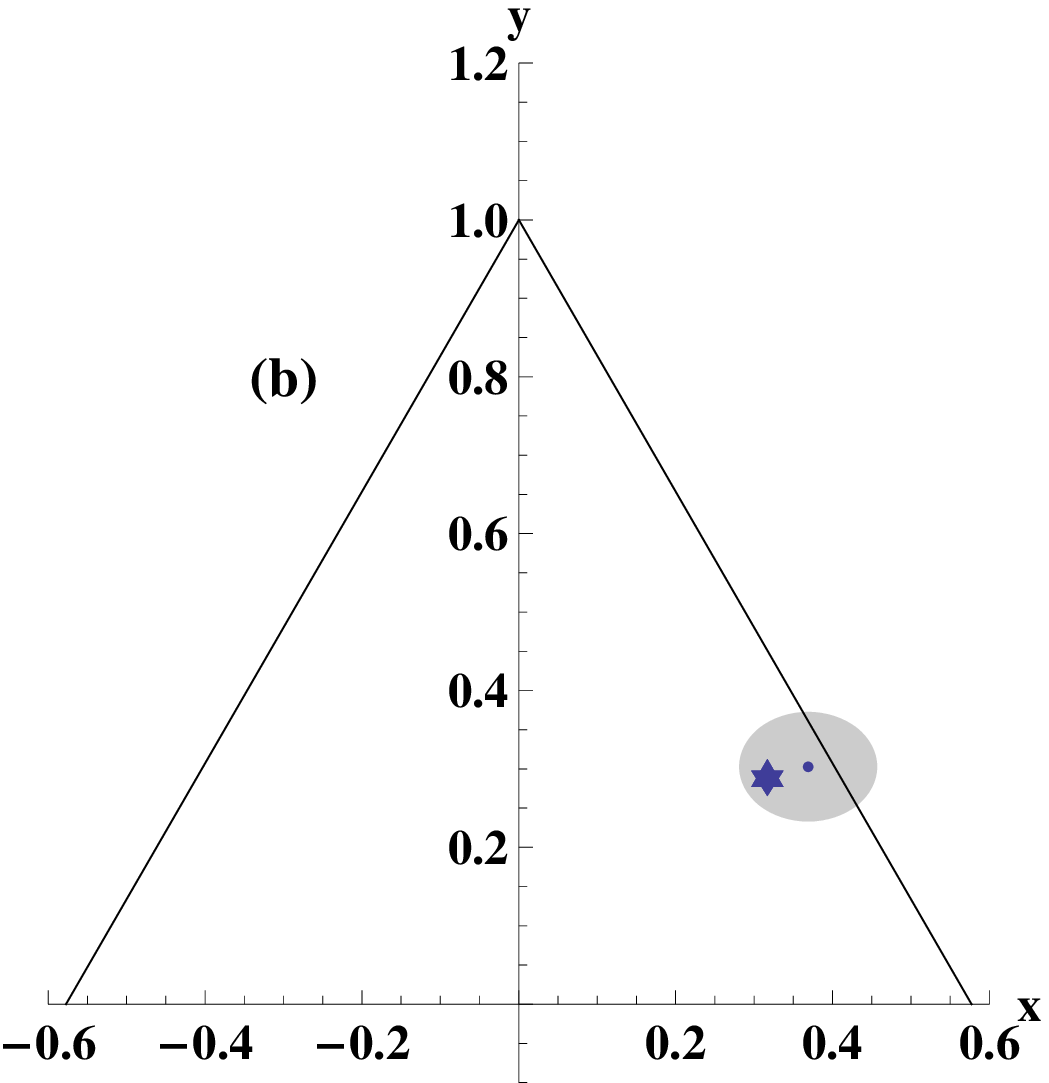,width=0.475\textwidth}\\
\end{tabular}
\caption{(a) is the joint plot of the TT, LT and LL correlation data  shown in 
figure~\ref{fig:rhos}. The coordinates $x$ and $y$ are defined in the text. The points are 
numbered 1 -- 7, the first one corresponding to the data bin in $\cos\theta_{W^{-}}$ at $-0.75$ and 
the last one to the bin at $\cos\theta_{W^{-}} = +0.9$. The curve inside the triangle plot is 
explained in the text.
\newline
(b)  shows the $(x,y)$ point obtained as the average of the points in~(a). The shaded area is one 
standard deviation  around the average $(x,y)$ point. The star symbol shows the corresponding 
point predicted by the Standard Model.} 
\label{fig:triangle}
\end{center}
\end{figure} 

\vspace{3mm}
\noindent 
Figure~\ref{fig:rhos} shows good agreement between the data and the Standard Model over the range 
of $\cos\theta_{W^{-}}$ where there are  sufficient data. An important aspect of these results is 
the interrelation between the three spin density correlations $\rho_{TT}$, $\rho_{LL}$ and
$\rho_{LT}$ determined in this analysis. These quantities can be displayed in one common plot due 
to the fact that they satisfy the condition~(\ref{eqn:unitarity}). Thus, they can be plotted  in a 
triangle plot, as shown in figure~\ref{fig:triangle}. Data can be plotted directly as indicated with 
arrows on 
the left side of figure~\ref{fig:triangle}(a) or by using  Cartesian coordinates  {\it x} and {\it 
y}, $$x=\frac{1}{\sqrt{3}}(\rho_{TT}-\rho_{LL})\ , \ y=\rho_{LT} \, .$$

\vspace{3mm}
\noindent
In figure~\ref{fig:triangle}(a) there are seven data points, each one corresponding to a different 
bin of $\cos\theta_{W^{-}}$. Error bars are also shown. The curve inside the triangle is the locus 
of points ({\it x,y}) calculated using CC03 diagrams. Each point on the curve corresponds to one 
particular value of $\cos\theta_{W^{-}}$. The point corresponding to $\cos\theta_{W^{-}} = -1$ is 
at the top vertex of the triangle. Further
points are distributed as indicated by the arrows along the curved line and the last point (for 
$\cos\theta_{W^{-}}=+1$) is back at the top vertex.
The solid part of the curve indicates the region where most of our data are located.

\vspace{3mm}
\noindent
Figure~\ref{fig:triangle}(b) shows the average values, $\bar{\rho}_{TT}$, $\bar{\rho}_{LT}$ and 
$\bar{\rho}_{LL}$, of the spin density matrix elements of Table~1, presented in a triangle plot.  
The shaded area is the one standard deviation region around the average point ({\it x,y}).  The 
star symbol is at the point predicted by the Standard Model using the CC03 diagrams.


\vspace{3mm}
\noindent 
It is clear that the TT correlation probability is large, the LT correlations are next in strength 
and the LL correlations are small. These are some of the most important features of the Standard 
Model predictions for the reaction $e^{-}e^{+}{\rightarrow}W^{-}W^{+}$.

\vspace{3mm}
\noindent
In spite of the limitations due to low statistics, these results show that the salient features of 
the Standard Model predictions for the $W - W$ polarisation correlations are compatible with our 
data.  This provides an additional test of the gauge theory relations between the 
$SU(2)_{L}\otimes SU(1)_{Y}$ couplings.

 
\subsection*{Acknowledgements}
\vskip 3 mm
We are greatly indebted to our technical 
collaborators, to the members of the CERN-SL Division for the excellent 
performance of the LEP collider, and to the funding agencies for their
support in building and operating the DELPHI detector.\\
We acknowledge in particular the support of \\
Austrian Federal Ministry of Education, Science and Culture,
GZ 616.364/2-III/2a/98, \\
FNRS--FWO, Flanders Institute to encourage scientific and technological 
research in the industry (IWT) and Belgian Federal Office for Scientific, 
Technical and Cultural affairs (OSTC), Belgium, \\
FINEP, CNPq, CAPES, FUJB and FAPERJ, Brazil, \\
Ministry of Education of the Czech Republic, project LC527, \\
Academy of Sciences of the Czech Republic, project AV0Z10100502, \\
Commission of the European Communities (DG XII), \\
Direction des Sciences de la Mati$\grave{\mbox{\rm e}}$re, CEA, France, \\
Bundesministerium f$\ddot{\mbox{\rm u}}$r Bildung, Wissenschaft, Forschung 
und Technologie, Germany,\\
General Secretariat for Research and Technology, Greece, \\
National Science Foundation (NWO) and Foundation for Research on Matter (FOM),
The Netherlands, \\
Norwegian Research Council,  \\
State Committee for Scientific Research, Poland, SPUB-M/CERN/PO3/DZ296/2000,
SPUB-M/CERN/PO3/DZ297/2000, 2P03B 104 19 and 2P03B 69 23(2002-2004),\\
FCT - Funda\c{c}\~ao para a Ci\^encia e Tecnologia, Portugal, \\
Vedecka grantova agentura MS SR, Slovakia, Nr. 95/5195/134, \\
Ministry of Science and Technology of the Republic of Slovenia, \\
CICYT, Spain, AEN99-0950 and AEN99-0761,  \\
The Swedish Research Council,      \\
The Science and Technology Facilities Council, UK, \\
Department of Energy, USA, DE-FG02-01ER41155, \\
EEC RTN contract HPRN-CT-00292-2002. \\


\end{document}